\documentclass[aps,prd,amsmath,amssymb,showpacs]{revtex4}
\usepackage{graphicx,psfrag}
\usepackage{bm}
\usepackage{epsfig}

\newcommand\fig[1]     {Fig.\,{\ref{#1}}}

\newcommand{\be}{\begin{equation}}
\newcommand{\ee}{\end{equation}}
\newcommand{\bea}{\begin{eqnarray}}
\newcommand{\eea}{\end{eqnarray}}
\newcommand{\hf}{\frac12}
\newcommand{\nn}{\nonumber\\}
\def\eq#1{(\ref{#1})}
\def\la{\langle}
\def\ra{\rangle}
\def\tr{{\mathrm{tr}}}
\def\Tr{{\mathrm{Tr}}}
\def\ord#1{{\cal O}\left(#1\right)}
\def\mr#1{{\mathrm{#1}}}
\def\v#1{{\bm{#1}}}

\def\u#1{{\underline{#1}}}
\def\ih{\frac{i}{\hbar}}
\def\hi{\frac{\hbar}{i}}

\def\cD{{\cal D}}
\def\cG{{\cal G}}

\def\yb{{\bar y}}
\def\Yb{{\bar Y}}
\def\gb{{\bar g}}

\def\gab{{\bar \gamma}}
\def\Gab{{\bar \Gamma}}
\def\Ib{{\bar I}}
\def\fd#1#2{\frac{\delta#1}{\delta#2}}
\def\fdd#1#2#3{\frac{\delta^2#1}{\delta#2\delta#3}}

\begin{document}

\title{Functional renormalization group for quantized anharmonic oscillator}

\author{S. Nagy}
\affiliation{Department of Theoretical Physics, University of Debrecen,
Debrecen, Hungary}

\author{K. Sailer}
\affiliation{Department of Theoretical Physics, University of Debrecen,
Debrecen, Hungary}

\date{\today}

\begin{abstract}
Functional renormalization group methods formulated in the real-time
formalism are applied to the $O(N)$ symmetric quantum anharmonic oscillator,
considered as a $0+1$ dimensional quantum field-theoric model,
in the next-to-leading order of the gradient expansion of the one- and two-particle 
irreducible effective action. The infrared scaling laws and the sensitivity-matrix analysis
show the existence of  only a single, symmetric   phase. The field-independent term
of the wavefunction renormalization turned out to be negligible, but its field-dependent
piece is noticeable. It is shown that the infrared limits of the running
couplings depend on the renormalization group  scheme used, when the perturbation
expansion in the bare quartic coupling is truncated keeping the terms up to the second order.
\end{abstract}


\maketitle

\section{Introduction}

Our main goal in this paper is to investigate the role of wave-function renormalization
in the quantum properties of the quartic anharmonic oscillator considering it as a
$0$(space)$+1$(time)-dimensional quantum field-theoric model, and applying various 
internal space renormalization group (RG)  schemes to the one- (1PI) and two-particle
irreducible (2PI)  effective actions in the truncated gradient expansion which goes beyond the local
potential approximation (LPA) with the incorporation of wave-function renormalization terms.
In that sense it is a continuation of the work presented in \cite{Hedd2004}. On the one hand, we
apply  a Callan-Symanzik (CS) type  internal space (IS) RG scheme to the 1PI effective action,
where an imaginary mass parameter serves as the control parameter of the RG evolution. On the other hand,
another IS RG scheme is applied to  the 2PI effective action, where the bare quartic
coupling $g_B$ represents the control parameter.
In both cases, the gradient expansion is used keeping the wavefunction renormalization
with, and the local potential is truncated beyond its quartic term.
The  results obtained in the framework of various RG schemes in the second-order
perturbation expansion in the coupling $g_B$ are also compared. 

The quantized anharmonic oscillator is an elementary toy model 
representing a building  block of more sophisticated and/or realistic models
finding several physical applications. Let us mention a few of them without pretending 
to completeness. The investigation of quantum properties
of nanomechanical oscillators has  became of interest in order to understand
the quantum behaviour of nanodevices \cite{Huang2009}. The statistical physical 
properties of chains of anharmonic oscillators, that of a one-dimensional Ginzburg-Landau 
system with quartic anharmonicities is of great interest in solid state physics.
2- and 3-dimensional aggregates of such chains describe anisotropic anharmonic solids
(e.g. ultrathin ferroelectric films). Putting such systems in inhomogeneous external
field enables one to manipulate individual monomers, making such films very attractive
for molecular electronics \cite{Bars2009}.  1-dimensional Ginzburg -Landau systems are very often 
investigated by means of the transfer matrix method in which the evaluation of the free energy of an $N+1$
dimensional classical system is equivalent to the evaluation of the ground state energy of
an $N$-dimensional quantum system, the so-called dual counterpart of the classical system.
In general  this dual problem is solved to understand its classical counterpart
\cite{Scal1972}, \cite{Krum1975}. Another field where the anharmonic oscillator model
  has got application is the study of the quantum physics of systems showing up chaotic
behaviour classically, ie. the study of quantum chaos \cite{Kowa2009}, \cite{Gevo2009}, \cite{Skok2009}.
Also the possibility of quantum effects induced chaos has recently been debated
 on the problem of the quantized Duffing oscillator \cite{Kapu2009}.

These widespreading possibilities of its applications explain the efforts to understand the
quantum physics of the anharmonic oscillator in all detail and to invent various theoretical tools 
for its investigation.  Already in the pioneering works  \cite{Bend1969},\cite{Bend1970} it
has been pointed out that the Rayleigh-Schr\"odinger  perturbation series for the ground state
energy of the quartic anharmonic oscillator diverges. Since then much efforts have been made
to overcome that problem and determine the eigenvalues and eigenfunctions of the
anharmonic oscillator by various methods like the strong-coupling expansion \cite{Bend1979},
the integration of operator differential equations \cite{Bend1989}, multi-scale perturbation
theory \cite{Bend1996}, variational methods \cite{Vlac1993},
iterative Bogoliubov transformations \cite{Jaur1992}, the eigenvalue moment method 
\cite{Hand1992}, the improved Hill-determinant method \cite{Chau1991}, the optimized
perturbation expansion \cite{Hats1997}, a particular iterative method based on the
generalized Bloch-equation \cite{Meis1997}, the auxiliary field method combined with loop expansion
\cite{Taro1999}, as well as the quantum computational method \cite{Sang2008}.
A class of non-perturbative approaches is based on various RG methods
\cite{Kuni1998,Yuka1998,Fras2002} including functional RG methods \cite{Kapo2000,Aoki2002},
\cite{Hedd2004} which treat the quantum mechanical problem as a  $0+1$ dimensional 
quantum field-theoric one.

The RG method has been basically developed for the purposes of
describing phase transitions and critical behaviour in statistical physics and very soon 
extended to investigate quantum field-theoric models. The RG strategy has been aimed
to monitor the Green functions of the elementary field variable when the quantum 
fluctuation modes are turned on gradually \cite{Wils1974}. While in the original version of
the RG method a finite fraction of modes are turned on in each blocking steps and one
generally relies on the perturbation expansion in bookkeeping the change of the dynamics,
the functional RG methods \cite{Wegn1973}-\cite{Morr1994}  turn on an infinitesimal 
fraction of modes in the subsequent blocking steps. Such an approach is accompanied with
the usage of infinitely many vertices. The manipulation of a high number of couplings
is carried out through their generating functional for which  one obtains an exact
RG equation looking like a one-loop equation. In order to solve that one has  to project it to
a suitably chosen functional subspace, e.g. by means of the truncated gradient expansion,
but that seems to be a physically better motivated approximation scheme than the truncation
of the perturbation expansion. The functional RG strategy has been developed for the 1PI
 effective action \cite{Wett1993}, \cite{Morr1994} with gliding momentum cut-off.
According to the original RG idea one organized the quantum fluctuations according
to their wavelengths or  momentum, ie. according to their properties in the external
space. A natural generalization came when such an ordering has achieved according to
their internal-space characteristics, the largeness of their amplitudes.
That step has been done by  developing the functional 
generalization of the CS equation \cite{Call1970}  where the mass 
parameter is evolved \cite{Alex2001} and the modes of the quantum fluctuations are
 turned on 
in the order of their increasing amplitudes. The CS RG scheme and the functional RG
 schemes
with gliding momentum cut-off give similar scaling laws in the leading order
in the UV scaling regime where the cutoff is the only scale parameter, but approaching the
IR scaling regime the cut-off dependence and the mass-dependence may differ due to
the occurring of additional physical scales. A further generalization of the CS RG
strategy is called IS RG \cite{Polo2005}, when any of the parameters of
the model can be used as control parameter and evolving it gradually one arrives at a
functional RG evolution equation for the 1PI effective action.
One should mention that  all IS RG schemes including the CS RG compare
various theories belonging to the gradually altering values of the control parameter,
while the RG schemes with gliding momentum cut-off provide true scaling laws for a 
given theory. So long the suppression term generating the IS RG evolution is quadratic,
the evolution equation contains one-loop integrals.
But it can be rather complicated when the suppression term is not quadratic in the
elementary field variable since the excitations described by the higher-loop integrals
appearing in the evolution equation have rather complicated structure in single-particle 
terms.  That problem can be overcome applying the internal space RG to the generating
functional of some composite operator of the elementary field, in which the suppression
term is again quadratic. In particular, in the case of a suppression term being quartic in the 
elementary field variable,  one may apply the RG method to the 
2PI effective action \cite{Polo2005}, the generating functional of 2PI 2-particle correlation
functions \cite{Domi1964,Haym1991}. That method has the  advantage that all expressions are  formally
ultraviolet (UV) finite until the generally UV  divergent loop integrals are made explicit 
making an Ansatz for the parametrization of the  one-particle Green function in whose terms
those are expressed.

As mentioned above, various functional RG methods have been recently used and proved 
to be powerful nonperturbative tools in treating the problem of the quantum anharmonic oscillator.
In \cite{Kapo2000} the effective average action RG scheme \cite{Wett1993} and  in \cite{Aoki2002}
the Wegner-Houghton (WH) RG method \cite{Wegn1973} have been applied to 
the one-dimensional quantum oscillator with quartic anharmonicity in order to determine the
energy of the ground state and that of the gap between the first excited state and the
ground state and both have been found to increase strictly monotonically  with increasing 
bare quartic coupling $g_B$. It has also been thoroughly discussed that even for a 
symmetric  double-well bare potential there exists  a unique ground state, so that the
quantum anharmonic oscillator exhibits only a single phase.  In the WH RG approach one
integrates out the high-frequency modes of quantum fluctuations in infinitesimal  steps 
above the gliding sharp momentum cut-off $k$. The method has the disadvantage that it 
disables one to go beyond the lowest order of the gradient
expansion, the so-called LPA which has also been used in
\cite{Aoki2002}.  Although the effective average action method allows for taking into
account wave-function renormalization terms in the gradient expansion, it has not been aimed in 
\cite{Kapo2000}.  In \cite{Hedd2004} the functional RG method has been applied to the
1PI effective action for single-particle quantum mechanics using the sharp gliding momentum 
cut-off as control parameter of the evolution. Truncating the expansion of the
1PI effective action at quartic terms,  RG flow equations for the 1PI two- and four-point
vertex functions have been derived. 
As an example of applications, the energy of the ground state  and that of the first excited 
state have been determined numerically.  The momentum-dependence of the proper self-energy
has been taken with, but the role of wave-function renormalization has not been
analysed directly. The authors found in accordance with the finding in 
\cite{Aoki2002} that the Lehman-expansion of the propagator is dominated by
the single pole corresponding to the first excited state that hints on the negligible role of 
the wavefunction renormalization.  Our purpose is to analyse the role of wavefunction
renormalization more thoroughly in terms of two different IS RG schemes, the
CS type one (CSi) with the imaginary mass as control parameter and another one  (ISg)
with the quartic coupling as control parameter.

First, we shall apply the CSi RG scheme to the $O(N)$-symmetric generalization of the one-dimensional
quartic oscillator considering the  oscillator coordinate 
$\u{q}=(q_1,\ldots, q_N)$ an $O(N)$ vector in the internal space. Due to 
$O(N)$ symmetry the anharmonic potential can only depend on powers of the $O(N)$ scalar
$\u{q}^2=\sum_{a=1}^N q_a^2=q_aq_a$. $O(N)$ symmetry couples any one-dimensional oscillator
$q_a$ $(a=1,2,\ldots, N)$ of the system to all of the others in a  rather particular manner even if we
restrict ourselves to a quartic potential, as we shall do throughout this work.
We shall derive the functional CSi RG equation for the 1PI effective action, $\Gamma[\u{q}]$
being the generating functional of the 1PI Green functions of the elementary field variable $\u{q}$.
In the right-hand side of the evolution equation the inverse matrix under the trace
shall be Neumann-expanded in its off-diagonal piece and the terms up to the quadratic ones
in the latter only  kept. The solution of that truncated evolution equation shall be looked for in the
next-to-leading order of the gradient expansion,  
including field-dependent wavefunction renormalization  besides the local potential.
In the case of quartic self-interaction the field-independent  wavefunction renormalization
emerges as a two-loop effect, therefore the way to generate it by  an essentially one-loop 
evolution equation opens  via including field-dependent wavefunction renormalization 
terms, as well. We shall take  with  only the quartic ones of those. The RG evolution of the
couplings and their IR values are then numerically investigated in the function of the
dimension $N$ of the oscillator. One should notice that the truncations used
may lead to erroneous results  for sufficiently large 
dimension $N$ and bare coupling $g_B$ \cite{Marg1998}.  We shall conclude that the
$O(N)$ symmetric anharmonic oscillator with any dimension $N$ exhibits a single phase 
and the IR couplings including those of the wavefunction renormalization depend smoothly 
of the dimension $N$.

At second,  we apply the ISg RG scheme to the second Legendre-transform of
the generating functional of connected Green-functions of the bi-local composite operator
$q_tq_{t'}$, to the 2PI effective action $\Gamma[G]$ that is the functional of the
single-particle  propagator $G$. According to our results obtained in the CSi RG scheme
no qualitative changes occur with increasing dimension $N$ in the trivial phase structure 
and the IR scaling laws of the anharmonic oscillator. Therefore we shall restrict our further
discussion to the simple anharmonic oscillator with a single degree of  freedom in the
internal space, ie. with $N=1$. The 2PI
effective action $\Gamma[G]$  takes its minimum at the propagator for the ground state
and its value at that minimum determines the energy of the ground state. Parameters of the
wave-function renormalization and the excitation energy of the first excited state  can be
read of from the propagator of the ground state. The renormalized  coupling of the quartic
term of the potential is determined via the second functional derivative of the effective 
action $\Gamma[G]$ at the ground state, the two-particle propagator of the vacuum in 
quantum field-theoric terms.  The RG evolution equation for the 2PI effective action shall be
derived  and solved with an Ansatz keeping the terms up to the  quadratic ones in the
2-particle propagator, called expansion in the 2-particle channel (E2PC).  Finally we shall
compare the results obtained in second order of the bare coupling $g_B$ by means of the CSi
and the ISg RG schemes, as well as with other results
taken from the literature. The dependence of the contributions of the order $\ord{g_B^2}$
on the RG schemes used shall be demonstrated.

The structure of the paper is as follows. Sect. \ref{sec:csrgn} contains the derivation of the 
CSi RG evolution equation for the 1PI effective action for the $O(N)$ symmetric anharmonic 
oscillator, the derivation of the coupled set of RG equations for the running couplings present
in the truncated gradient expansion of the
effective action, and the comparison of the numerical  solutions obtained in the independent
mode approximation (IMA), the LPA, and the approximation including wavefunction
renormalization (AWF).  The existence of only a single phase is stressed by discussing the 
sensitivity of the IR couplings to their bare values, as well.  Sect. \ref{sec:isrg} starts with 
the derivation of the RG evolution equation for the 2PI effective action for the one-component
$(N=1)$ quartic anharmonic oscillator,  that is followed by the solutions first in the IMA,
and afterwards  in the  truncated E2PC. Finally, the results for the various couplings are
determined keeping the terms up to the order $\ord{g_B^2}$  and compared with
the results obtained in the CSi RG scheme in second order perturbation expansion, as described in
Appendix \ref{a:cspe}, with the WH RG scheme in the same approximation shortly outlined in Appendix \ref{a:wheuc}
and the well-known results of the Rayleigh-Schr\"odinger perturbation expansion (RS PE).  
In Sect. \ref{sec:con} the conclusions are drawn.
The details of the derivation of CSi RG evolution equations are given in Appendix \ref{a:nemex},
the loop integrals needed in both schemes are given in Appendix \ref{a:loopCS2}.
Appendix \ref{a:wheuc} presents a short derivation of the WH RG equation for a one-dimensional
quantum mechanical system and its application to the anharmonic oscillator in the second order
of the perturbation expansion.

\section{CSi RG for the N-dimensional oscillator}\label{sec:csrgn}

\subsection{RG equations}

In the Euclidean formulation of the quantum field-theoric models the CS
RG scheme is realized by evolving the mass as a control parameter  from above the UV
momentum cut-off $\Lambda$ towards its physical value continuously.  The evolution accounts for the
quantum fluctuations  gradually according to their increasing amplitude. Here we realize
the similar RG procedure in the real-time formulation by introducing an evolving
imaginary mass as control parameter which vanishes in the IR limit. Euclidean flow 
equations being easier to manage numerically are
then obtained by analytic continuation of the control parameter. The advantage of the
real-time approach is that the path integral remains well-defined even for a quantum system with 
double-well potential, ie. for negative values of the bare parameter $\omega_0^2$. 
Nevertheless one does not expect any phase  transition with analytic continuation of the
$\omega_0^2$ parameter to the negative real axis, since there emerges no spontaneous 
symmetry breaking in quantum mechanics. The large amplitude quantum fluctuations 
fill up the potential wells and make it convex in the IR limit (c.f. with the findings in Refs.
\cite{Kapo2000,Aoki2002}).

The basic idea is that we introduce an imaginary part $\hf i\mu^2\int_t \u{q}_t^2$
into the quadratic term of the bare action $S_B[\u{q}]$, in order to suppress the quantum fluctuations
in the path integral for $\mu^2 \approx \Lambda^2$. The real control parameter $\mu^2$ is
then continuously decreased to zero.  The imaginary quadratic part ensures the convergence
of the path integral during the whole evolution.  The generator functional of the connected
Green-functions of the elementary `field variable' $\u{q}_t$, the time-dependent coordinate of the
linear $O(N)$ symmetric anharmonic oscillator is given as
\bea
S_B[\u{q}]&=& \int_t \biggl( \hf {\dot {\u{q}}}^2 - \hf (\omega_0^2-i\mu^2) \u{q}^2 
-\frac{g_B}{24}({\u{q}}^2)^2 \biggr).
\eea
For $\mu^2=\mu^2_B
\approx \Lambda^2$, the quantum fluctuations are frozen and they can be gradually
`melted out' when the control parameter is decreased towards $\mu^2\to 0$. For the 
sake of simplicity the  mass $m$ of the oscillator is set to unity. 
Throughout the paper we shall use the notations and conventions
$\int_t=\int_{-T/2}^{T/2} dt$, $(T\to \infty)$, $\int_\omega
=\int_{-\Lambda}^\Lambda \frac{d\omega}{2\pi}$, $(\Lambda\to \infty)$,
$\sum_{a=1}^N\int_t f_{a,t} g_{a,t}= \u{f}\cdot \u{g}$,
$f_{a,\omega}= \int_t e^{i\omega t}f_{a,t}$, $\Ib_{(a,t),(b,t')}=\delta_{a,b}\delta_{t,t'}
=\delta_{a,b}\delta (t-t')$,
 $\Ib_{(a,\omega),(b,\omega')}=\delta_{a,b}\delta_{\omega+\omega',0}=
\delta_{a,b}\int_{t,t'}e^{i\omega t+i\omega't'} \delta_{t,t'}$,
 $\delta_{\omega=0,0}=T$, $\tr A=\sum_{a=1}^N\int_{t,t'}\delta_{t,t'}A_{(a,t),(a,t')}
=T\sum_{a=1}^N\int_\omega A_{(a,\omega),(a,-\omega)}$, and identical Latin indices
denote summation like $f_ag_a=\sum_{a=1}^Nf_ag_a$.
The generating functional of the connected Green-functions of the elementary variable $\u{q}_t$
is defined via the path integral
\bea
   e^{\ih W[\u{j}] } &=& \int {\cal D}\u{q} e^{\ih S_B[\u{q}]+ \ih \u{j}\cdot \u{q} }
\eea
in the presence of the time-local external source $\u{j}_t$. The generating functional of the 1PI
Green functions, the 1PI effective action as the functional of the ground-state expectation
value of the coordinate,  $  q_{a,t}= \delta W[j]/\delta j_{a,t}$ is defined by the
Legendre-transform
\be
{\v{-}}\Gamma[\u{q}]=-W[\u{j}] + \u{j}\cdot \u{q}
\ee
for which $j_{a,t}= \v{-}\delta \Gamma[\u{q}]/\delta q_{a,t}$, and $\Gamma^{(2)}\cdot W^{(2)} =\v{-}\Ib$
where $W^{(2)}$ and $\Gamma^{(2)}$ stand for the second functional derivatives
of the functionals $W[\u{j}]$ and $\Gamma[\u{q}]$, respectively. 
In order to find the CS-type functional evolution equation, we evaluate
the partial derivative of the effective action with respect to the control parameter $\mu^2$, 
\bea
\partial_{\mu^2} \Gamma[\u{q}]&=&\v{+}\partial_{\mu^2} W[\u{j}] =\v{+}e^{-\ih W[\u{j}]} \int {\cal D} \u{q} 
\int_t \hf i {\u{q}}^2 e^{\ih S[\u{q}] +\ih \u{j}\u{q} }
=\v{+}e^{-\ih W[\u{j}]} \hf i \int_t \fdd{}{\ih j_{a,t}}{\ih j_{a,t}} e^{\ih W[\u{j}]}
 \nn
&=&\v{+}\hf i \int_t \biggl(  \hi \fdd{W[\u{j}]}{j_{a,t}}{j_{a,t}} + \fd{W[\u{j}]}{j_{a,t}}
 \fd{W[\u{j}]}{j_{a,t}}\biggr)
= \v{+}\frac{\hbar}{2} \biggl(\v{-} \tr \Gamma^{(2)~-1} + \ih
  \u{q}\cdot \u{q} \biggr).
\eea
Now we introduce the reduced effective action $\Gab[\u{q}]$ with the relation
$\Gamma[\u{q}]=\Gab[\u{q}] \v{+} \hf i\mu^2 \u{q}\cdot \u{q}$ in terms of which the
evolution equation takes the following form,
\bea\label{eveq}
\partial_{\mu^2} \Gab[\u{q}] &=& -\frac{\hbar}{2} \tr ( \Gab^{(2)}[\u{q}] \v{+}i\mu^2 )^{-1} .
\eea

We shall look for the solution of the evolution equation in the truncated gradient expansion
by making use of the Ansatz
\bea\label{ans}
  \Gab[\u{q}]&=& -T\gamma_{\mu^2}
 +\hf \int_\omega ( x_{\mu^2} \omega^2-\Omega_{\mu^2}^2) \u{q}_\omega \u{q}_{-\omega} 
-  \int_{\omega_1,\ldots,\omega_4} v_{\mu^2}(\omega_1,\omega_2)
 \delta_{\omega_1+\ldots+\omega_4,0}  q_{a,\omega_1}q_{a,\omega_2} q_{b,\omega_3}q_{b,\omega_4}
  \eea
with $x_{\mu^2}$ for field-independent wave-function renormalization and 
\be
  v_{\mu^2}(\omega_1,\omega_2)= 
\frac{1}{2} {\bar y}_{\mu^2} \omega_1\omega_2
   + \frac{1}{4} {\bar Y}_{\mu^2} (\omega_1^2+\omega_2^2)
  + \frac{g}{24} = v_{\mu^2}(\omega_2,\omega_1)= 
 v_{\mu^2}(-\omega_1,-\omega_2),
\ee    
where the first two terms stand for the field-dependent, quadratic wavefunction renormalization,
\be\label{fdwfr}
 - \int_t \biggl(\hf {\bar y}_{\mu^2}  {\u{q}}^2 {\dot {\u{q}}}^2
    + \hf {\bar Y}_{\mu^2} {\u{q}}^2 ({\ddot {\u{q}}}\u{q} )\biggr) ,
\ee
and the third term for the quartic self-interaction. 
Integrating by parts one finds
\bea\label{ident}
   - \int_t {\ddot q}_{a,t}q_{a,t} q_{b,t}q_{b,t}
&=& \int_t [{\dot q}_{a,t}{\dot q}_{a,t}q_{b,t}q_{b,t} + 2 {\dot q}_{a,t}q_{a,t} {\dot q}_{b,t}q_{b,t}],
\eea
so that one can perform the replacement $\omega_1^2 \Leftrightarrow  (\omega_1\omega_2 +2 \omega_1\omega_3) $
in the kernel of the last integral term in the right-hand side of Eq. \eq{ans} and write
\bea\label{fdwfr2}
  - \int_t \biggl(\hf {\bar y}_{\mu^2}  {\u{q}}^2 {\dot {\u{q}}}^2
    + \hf {\bar Y}_{\mu^2} {\u{q}}^2 ({\ddot {\u{q}}}\u{q} )\biggr)&=&
 - \int_t \biggl( \hf y_{\mu^2}  {\u{q}}^2 {\dot {\u{q}}}^2
 + \hf Y_{\mu^2}({\dot {\u{q}}}\u{q})^2 \biggr)
\eea
$  y_{\mu^2}= {\bar y}_{\mu^2}-{\bar Y}_{\mu^2}$ and
$  Y_{\mu^2}= -2 {\bar Y}_{\mu^2}$.

Let us separate off the diagonal piece ${\bar G}^{-1}$ and the off-diagonal piece $A$ of the matrix
\be
 \Gab^{(2)}[\u{q}]\v{+} i\mu^2 = {\bar G}^{-1}+A,
\ee
where $  {\bar G}_{(c,\omega),(d,\omega')}=G_\omega \delta_{c,d}\delta_{\omega+\omega',0}$
with $G_\omega=[x_{\mu^2} \omega^2  -\Omega_{\mu^2}^2\v{+} i\mu^2]^{-1}$ and
\bea
  A_{(c,\omega),(d,\omega')} &=& 
-\fdd{}{q_{c,{\bf -\omega}}}{q_{d,{\bf -\omega'}}} \int_{\omega_1,\ldots,\omega_4}
\delta_{\omega_1+\ldots+\omega_4,0} v_{\mu^2}(\omega_3,\omega_4)
 q_{a,\omega_1}q_{a,\omega_2}q_{b,\omega_3}q_{b,\omega_4}
=
  \delta_{c,d}B_{\omega,\omega'}+ {\bar C}_{(c,\omega),(d,\omega')}
\eea
with
\bea
   B_{\omega,\omega'} &=& -2  \int_{\omega_1,\omega_2}\delta_{-\omega-\omega'+\omega_1+\omega_2,0} 
[v_{\mu^2}(\omega_1,\omega_2)+ v_{\mu^2}(\omega,\omega')]
q_{a,\omega_1}q_{a,\omega_2} =B_{\omega',\omega},\nn
 {\bar C}_{(c,\omega),(d,\omega')}&=&
-4 \int_{\omega_1,\omega_2}\delta_{-\omega-\omega'+\omega_1+\omega_2,0} 
[v_{\mu^2}(\omega_1,-\omega)+v_{\mu^2}(\omega_2,-\omega') ]
q_{c,\omega_1}q_{d,\omega_2}= {\bar C}_{(d,\omega'),(c,\omega)}.
\eea
As to the next we Neumann-expand  the inverse matrix in the right-hand side of the
evolution equation \eq{eveq}, keeping the terms up to the second order in the off-diagonal piece,
\bea\label{neuexp}
-\frac{\hbar}{2}\tr[ {\bar G}^{-1}+A]^{-1}
  &=&-\frac{\hbar}{2} \tr [{\bar G}-{\bar G}A{\bar G}+{\bar G}A{\bar G}A{\bar G} -+\ldots]\nn
&=&
-\frac{ N\hbar}{2}\int_\omega G_\omega 
+\frac{\hbar}{2} \int_{\omega} G_\omega^2 A_{(a,\omega),(a,-\omega)}
-\frac{\hbar}{2}\int_{\omega,\omega'} G_\omega^2 A_{(c,\omega),(d,\omega')}G_{\omega'}A_{(d,-\omega'),(c, -\omega)}-+\ldots 
\eea

\begin{figure}[thb]
\psfrag{dm2}{\Large $\partial_{\mu^2}=$}
\psfrag{gg}{\Large $g=$}
\psfrag{yy}{\Large $y=$}
\psfrag{intqdw}{\Large $\int qd\omega=$}
\psfrag{intwqdw}{\Large $\int \omega qd\omega=$}
\psfrag{AA}{\Large $A=$}
\psfrag{dm2G}{\Large $\partial_{\mu^2}\Gamma=$}
\includegraphics[width=12cm]{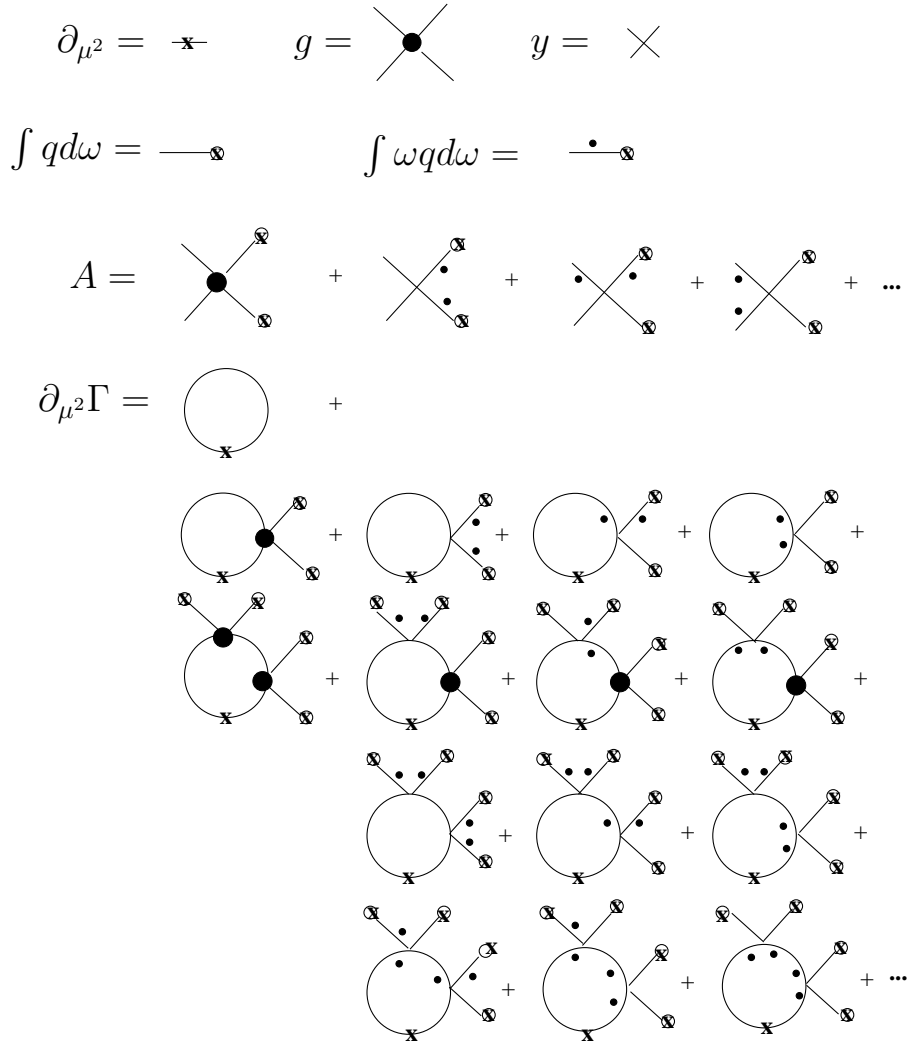}
\caption{Feynman diagrams for the vertices and the right-hand side of the evolution 
equation in CSi RG
for the $N=1$ dimensional oscillator.
\label{fig:cspargam}}
\end{figure}

It is worthwhile noticing here that for $N=1$ the terms for field-dependent wavefunction
renormalization given in Eqs. \eq{fdwfr} or \eq{fdwfr2}  merge into a single one,
\be
  - \int_{\omega_1,\ldots,\omega_4}
\delta_{\omega_1+\ldots+\omega_4,0} \hf \Upsilon_{\mu^2} \omega_1\omega_2
 q_{\omega_1}q_{\omega_2}q_{\omega_3}q_{\omega_4}
\ee
with $\Upsilon_{\mu^2}=\yb_{\mu^2}-3\Yb_{\mu^2}$. It is  illuminating to give the graphical representation
of the first few terms of the Neumann-expansion in the functional  CSi
equation \eq{eveq} for the case $N=1$, shown in Fig. \ref{fig:cspargam}.

The details of the evaluation of the integrals in the right-hand side of Eq. \eq{eveq} are given in Appendix 
\ref{neumann}.  In addition to the truncation of the Neumann-expansion we have 
expanded each term of the Neumann-series in powers of the Fourier-transform $q_{a,\omega}$ of the
field variable  as well as in powers of its frequency index $\omega$
and neglected the higher-order terms   absent in  Ansatz
\eq{ans}. In order to perform the expansion in the frequency (i.e. in the time-derivative
of the field variable) we used the expansion of the propagator,
\bea\label{exprop}
  G_{\omega+ \alpha}&=& 
 G_\omega- 2x_{\mu^2} \omega G_\omega^2  \alpha
-  [1 - 4 x_{\mu^2} \omega^2G_\omega] x_{\mu^2} G_\omega^2  
\alpha^2+\ord{\alpha^3}.
\eea
Furthermore, we have to make use of  identity \eq{ident} in order to transform all derivative terms into
one of the forms present in Ansatz \eq{ans}.
The truncation the Neumann-expansion and that of the gradient expansion may lead to
erroneous results in the IR scaling regime for sufficiently strong coupling $g_B$ and/or large number $N$ of
dimensions. The steps described above enable one to express
the trace in the right-hand side of the evolution equation \eq{eveq} in terms of the loop integrals
\be\label{csrealoop}
  I_{n,s}(\mu^2, \Omega^2_{\mu^2})=
\hbar \int_{-\Lambda}^\Lambda \frac{d\omega}{2\pi}   G_\omega^n\omega^s
\ee
and find the evolution equation
\bea\label{eveq1}
&&
 -T\partial_{\mu^2}\gamma_{\mu^2}
+ \hf \int_{\omega_1} ( \partial_{\mu^2} x_{\mu^2} \omega_1^2
-\partial_{\mu^2}\Omega_{\mu^2}^2 )q_{a,\omega_1}q_{a,-\omega_1}\nn
&&
-\int_{\omega_1,\ldots,\omega_4} \delta_{\omega_1+\ldots+\omega_4,0}
\biggl( \hf \partial_{\mu^2} \yb_{\mu^2} \omega_1\omega_2
  + \hf \partial_{\mu^2} \Yb_{\mu^2} \omega_1^2
 + \frac{1}{24}\partial_{\mu^2} g
\biggr)q_{a,\omega_1}q_{a,\omega_2}q_{b,\omega_3}q_{b,\omega_4}    \nn
= &&
-\frac{TN}{2}I_{1,0}
+\hf [N\yb_{\mu^2}-(N+2)\Yb_{\mu^2} ] I_{2,0} \int_{\omega_1}\omega_1^2 
q_{a,\omega_1}q_{a,-\omega_1}  \nn
&&
+\hf  \biggl( [N\yb_{\mu^2}-(N+2)\Yb_{\mu^2}] I_{2,2}
 - \frac{(N+2) g_{\mu^2}}{6} I_{2,0}
\biggr)\int_{\omega_1}  q_{a,\omega_1}q_{a,-\omega_1}
+T_1+T_2+T_3
\eea
where the purely quartic terms $T_i$ $(i=1,2,3)$ are given by Eqs. \eq{t1}-\eq{t3}.
In the limit $\Lambda\to \infty$ the loop-integrals can be taken analytically, and all expressed in terms of a few ones, 
$I_{n,0}$ with $n=1,2,3,4$, (see Appendix \ref{a:loopCS2}). Comparing the coefficients of the
corresponding terms of the gradient expansion in both sides of Eq. \eq{eveq1}, one arrives at
the RG evolution equations for the couplings in the our approximation, called previously AWF,
\bea\label{ga}
 \partial_{\mu^2}\gamma
&=&
\frac{N}{2}I_{1,0},
\eea
\bea\label{om2}
\partial_{\mu^2}\Omega^2 &=&
-  [N\yb-(N  +2)\Yb] I_{2,2}
 +\frac{(N+2) g}{6} I_{2,0},
\eea
\bea\label{g}
 \partial_{\mu^2}g &=&
12[ N\yb^2+ 2(N+2)\yb\Yb +(N+8)\Yb^2 ]I_{3,4}
-4g[ (N+2)\yb -(N+8)\Yb]I_{3,2}
+ \frac{(N+8)g^2}{3}I_{3,0},
\eea
\bea\label{x}
 \partial_{\mu^2} x &=&
[N\yb-(N+2)\Yb] I_{2,0},
\eea
\bea\label{yb}
\partial_{\mu^2} \yb &=&
8x^2[ N\yb^2-2(N+2)\yb\Yb+(N+2)\Yb^2]I_{5,6}
-\frac{8(N+2)g}{3} x^2(\yb-\Yb) I_{5,4}
+\frac{2(N+2)g^2}{9}x^2 I_{5,2}\nn
&&
-10x[N\yb^2-2(N+2)\yb\Yb+(N+2)\Yb^2]I_{4,4}
+2(N+2)g x (\yb-\Yb)I_{4,2}
-\frac{(N+2)g^2}{18}x I_{4,0}\nn
&&
+4[ -2\yb^2- (N+2)\yb\Yb+ (N+2)\Yb^2]I_{3,2}
+\frac{(N+2)g}{3}(\yb+\Yb)I_{3,0},
\eea
\bea\label{Yb}
\partial_{\mu^2}\Yb &=&
8x^2[ N\yb^2-2(N+2)\yb\Yb+(N+6)\Yb^2]I_{5,6}
-\frac{8g}{3}x^2 [(N+2)\yb-(N+6)\Yb]I_{5,4}
+\frac{2(N+6)g^2}{9}x^2 I_{5,2}\nn
&&
-10x [N\yb^2 -2(N+2)\yb\Yb +(N+6)\Yb^2]I_{4,4}
+2gx[(N+2)\yb -(N+6)\Yb]I_{4,2}
-\frac{(N+6)g^2}{18}xI_{4,0}\nn
&&
+[2N\yb^2-8(N+3)\yb\Yb+ (6N+40)\Yb^2]I_{3,2}
-\frac{2g}{3}[ 2\yb -(N+7)\Yb]I_{3,0}.
\eea
For the sake of simplicity, we have suppressed the lower index $\mu^2$ of the running couplings.
The loop-integrals occurring in these equations are UV finite except of $I_{1,0}$, so that
the IR limits of all couplings can be determined  for $\Lambda \to \infty$ with exception of
the constant term of the 1PI effective action determining the energy of the ground state. 
The $x$-dependence of the right-hand sides of the evolution equations is partly explicit,
partly implicit occurring  via the propagators in the loop-integrals. The evolution of the
field-independent wavefunction renormalization $x$ is generated via the evolution
of the field-dependent wavefunction renormalization parametrized by $\yb$ and $\Yb$
(c.f. Eq. \eq{x}) because those start to evolve already at the UV scales due to the terms
containing the loop-integrals $I_{5,2}$ and $I_{4,0}$ in the right-hand sides of
Eqs. \eq{yb} and \eq{Yb} even for vanishing bare values $\yb_B=\Yb_B=0$.
The evolution equations for the LPA can easily be obtained by setting $x=1$ and
$\yb=\Yb=0$ identically. It is worthwhile noticing that the analytic continuation of the RG equations to Euclidean 
space via the replacement $\v{-i\mu^2}\to \lambda$ for real $\lambda$ decreasing from
$\lambda_B\approx\Lambda^2$ towards the IR limit $\lambda\to 0$ is rather suitable
for the  numerical treatment since the Euclidean RG flow does not develop imaginary parts
for the couplings.

For $N=1$ only the single coupling $\Upsilon=\yb-3\Yb$ for field-dependent wavefunction
renormalization occurs and  the RG equations \eq{ga}-\eq{Yb} reduce to the following ones,
\bea\label{gan1}
 \partial_{\mu^2}\gamma &=&\frac{1}{2}I_{1,0},
\eea
\bea\label{om2n1}
\partial_{\mu^2}\Omega^2 &=&-  \Upsilon I_{2,2}
+\hf g I_{2,0},
\eea
\bea\label{gn1}
 \partial_{\mu^2}g &=&
 12 \Upsilon^2 I_{3,4}
- 12g\Upsilon I_{3,2}
+ 3 g^2 I_{3,0},
\eea
\bea\label{xn1}
 \partial_{\mu^2} x &=& \Upsilon I_{2,0},
\eea
\bea\label{upsi}
\partial_{\mu^2} \Upsilon &=&
 -16x^2 \Upsilon^2 I_{5,6}
+16g    x^2 \Upsilon I_{5,4}
 -4 g^2 x^2 I_{5,2}
+20 \Upsilon^2 x I_{4,4}
-12g  x \Upsilon I_{4,2}
+g^2 xI_{4,0}
-14 \Upsilon^2 I_{3,2}
+5 g \Upsilon I_{3,0}.
\eea
We see now for $N=1$ that the parameter $x$ starts to evolve in the UV scaling  regime 
because of the parameter $\Upsilon$ does due to the nonvanishing terms 
$ -4 g^2 x^2 I_{5,2}+g^2 xI_{4,0}$ in the right-hand side of  Eq. \eq{upsi}.

\subsection{Solution of the CSi RG equations} 

\subsubsection{IMA}

The IMA corresponds to the replacement of the running couplings by their bare
values  $\Omega^2_B=\omega_0^2$, $g_B$, $x_B=1$, $\yb_B=\Yb_B=0$
into the right-hand sides of the evolution equations \eq{gan1}-\eq{upsi}.
Making use of the loop integrals of Appendix \ref{a:loopCS2} in the limit $\Lambda\to \infty$
and the analytic continuation to the Euclidean space via the replacement
$\lambda=\v{-i\mu^2}$, one can easily integrate Eqs.  \eq{gan1}-\eq{upsi} from the initial
scale $\lambda_B\to \infty $ down to the gliding scale $\lambda$. The UV divergence
of the ground-state energy can be removed by the choice $\gamma_{B}=  \hf N\hbar[\omega_0^2+\lambda_B]^\hf $.
The  UV scaling laws obtained in that manner can then be extrapolated to the IR limit $\lambda\to 0$:
\bea\label{gaima}
&&\gamma_0= \hf N\hbar \omega_0,~~
\Omega^2_0 = 
\omega_0^2 \biggl[ 1 +  \frac{4(N+2) \xi }{3} \biggr],~~
g_0 =  g_B\biggl[ 1- \frac{ 2 (N+8)\xi }{3}\biggr],\nn
&&  x_0 =1,~~
\hbar\yb_0 = -\frac{4(N+2)\omega_0 \xi^2}{9},~~
\hbar \Yb_0=- \frac{ 4 (N+6)\omega_0 \xi^2}{9},
\eea
where the dimensionless parameter $\xi= \frac{g_B\hbar}{16\omega_0^3}$
of the Rayleigh-Schr\"odinger perturbation expansion \cite{Land1965}  has been introduced.
We see that the one-loop corrections of the couplings depend linearly on the dimension $N$.
The energy of the ground state  $E_0=\gamma_0=\hf N\hbar \omega_0$ is
just the sum of the zero-point energies of the number $N$ of independent harmonic 
oscillators in this approximation, the one-loop result. The energy gap between the first 
excited state and the ground state, i.e. in field-theoric terms the minimum energy of the 
one-particle excitations, 
$
E_1-E_0=\hbar \sqrt{ \Omega_0^2/x_0}=
\hbar\omega_0 \sqrt{  1+  4(N+2) (\xi/3)   }$
increases with increasing bare coupling strength $g_B$ and number of dimension $N$.
The extrapolated IR value of the quartic coupling $g_0$  decreases with its increasing $g_B$ and $N$  in the IMA.
Although there occurs a field-dependent wavefunction renormalization of the order $\ord{g_B^2}$ in the IMA, 
the field-independent wavefunction renormalization is absent since it is a two-loop effect
in the lowest  nonvanishing order of the loop expansion. The beta-functions do not change sign, so that
one can conclude that the extrapolation of the UV scaling laws to the IR region seems to 
predict only a single phase of the quantum oscillator independently of its dimension $N$.
However, restricting ourselves to that approximation, we would get in trouble with the change of sign of $g_0$
for sufficiently large $g_B$ and $N$, as well as with a continuation of these formulae to negative values of
$\omega_0^2$ corresponding to a symmetric double-well bare potential.

\subsubsection{Numerical integration of the CSi RG equations}

\begin{figure}[ht] 
\begin{center} 
\epsfig{file=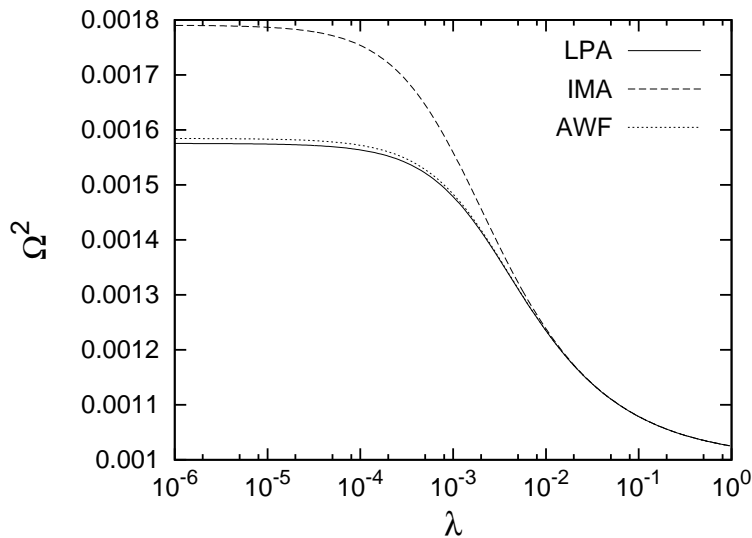,width=7cm,angle=-90}
\caption{\label{fig:appr}
The flow of the frequency parameter $\Omega^2$  in different approximations,
 for $\Omega^2_B=0.00102$, $g_B=0.0001$, $x_B=1$, $\yb_B=\Yb_B=0$,
and $N=1$.
} 
\end{center}
\end{figure}

\begin{figure}[ht] 
\begin{center} 
\epsfig{file=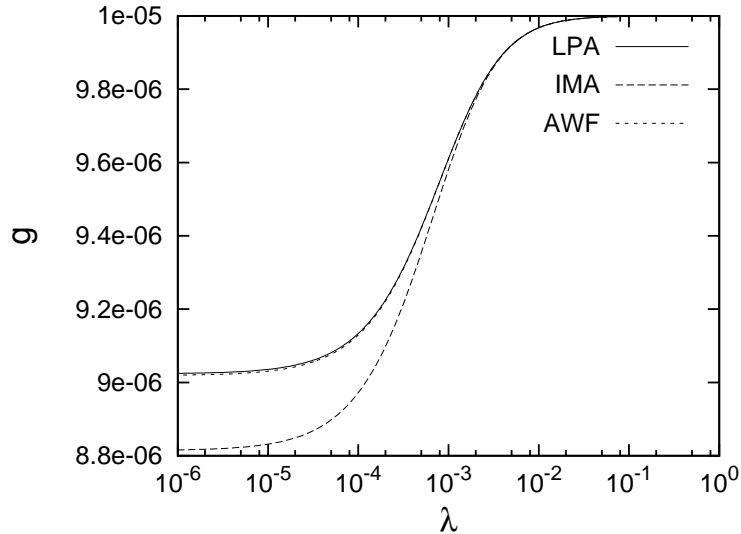,width=7cm,angle=-90}
\caption{\label{fig:apprg}
The flow of the quartic coupling $g$ in different approximations, for 
$\Omega^2_B=0.00102$, $g_B=0.00001$, $x_B=1$, $\yb_B=\Yb_B=0$, and $N=1$.
} 
\end{center}
\end{figure}

The numerical integration of the RG equations \eq{om2}-\eq{Yb} obtained in the CSi RG scheme in the AWF and 
analytically continued to the Euclidean space
has been performed by means of a fourth-order Runge-Kutta algorithm.
In Figs. \ref{fig:appr} and \ref{fig:apprg} the RG flow obtained in AWF
is compared to  RG flows obtained in the IMA and the LPA. (The latter is obtained by setting
$x=1,~\yb=\Yb=0$ identically.) 
One can see, that the flows in the various approximations
are  qualitatively  the same, i.e. after the UV scaling region both the
dimensionful frequency $\Omega^2$ and the quartic coupling $g$  saturate at 
constant  values in the IR region.  The UV scaling is described 
properly by all approximations, but they give different values in the IR limit.
The IMA corresponding to the one-loop approximation gives the largest value for $\Omega^2$ and
the smallest one for $g$. More reliable
results are obtained in the LPA and AWF which resum infinitely many Feynman diagrams.
Both of the Figs. \ref{fig:appr} and \ref{fig:apprg} clearly show that the 
inclusion of the wavefunction renormalization modifies only slightly the evolution as 
compared to  the LPA. This is a consequence of the fact that the pole of the propagator, 
corresponding to the first excited state of the oscillator dominates in its Lehman-expansion,
as discussed in \cite{Aoki2002}.

\begin{figure}[ht] 
\begin{center} 
\epsfig{file=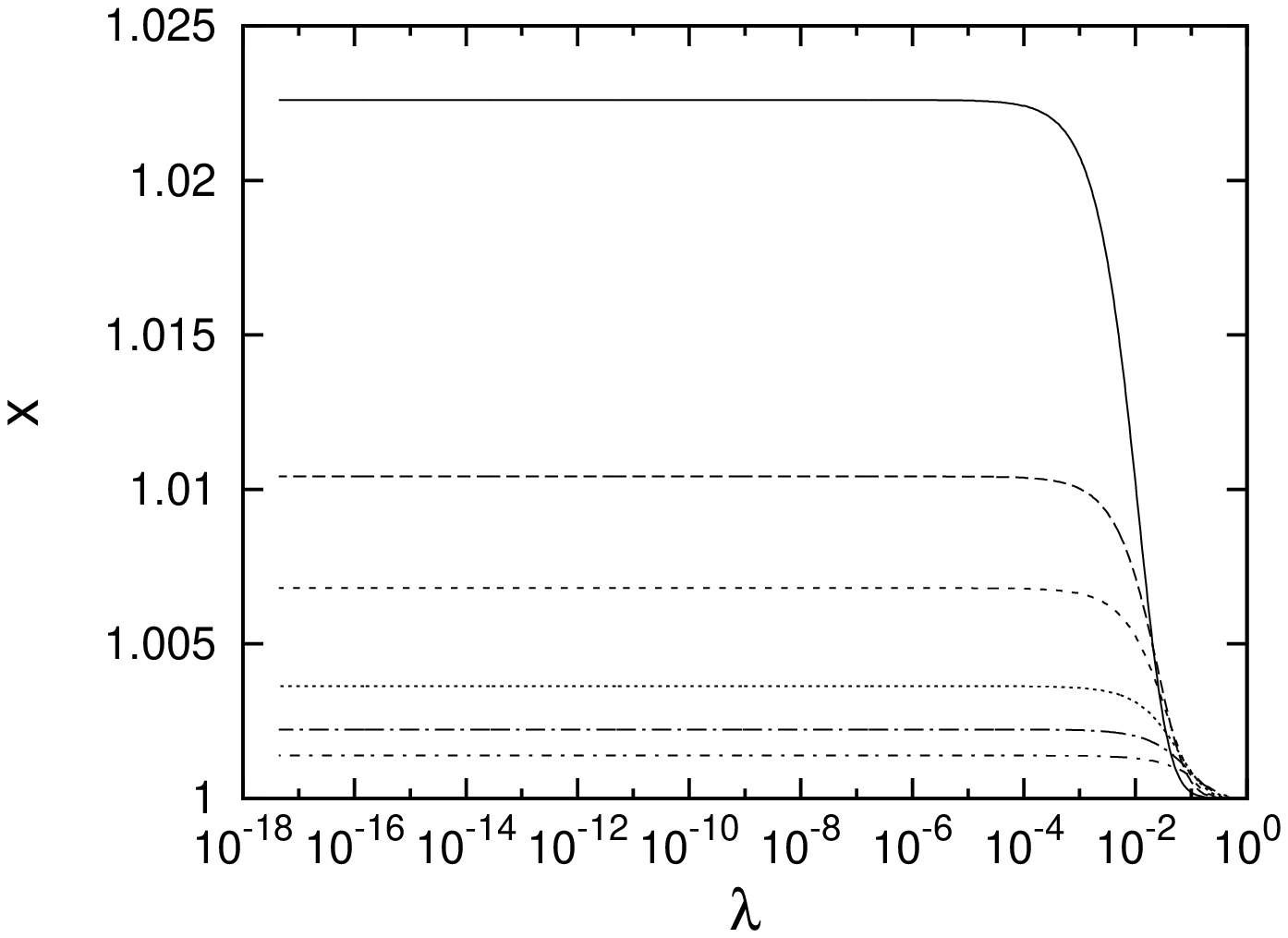,width=7cm,angle=-90}
\caption{\label{fig:nwfx}
The scale dependence of $x$ for $\Omega^2_B=0.001,~g_B=0.01$,
and $\yb=\Yb=0$.
The order of the  curves from the top  correspond to the various values  $N=2,10,20,50,100,200$, 
respectively.
} 
\end{center}
\end{figure}

\begin{figure}[ht] 
\begin{center} 
\epsfig{file=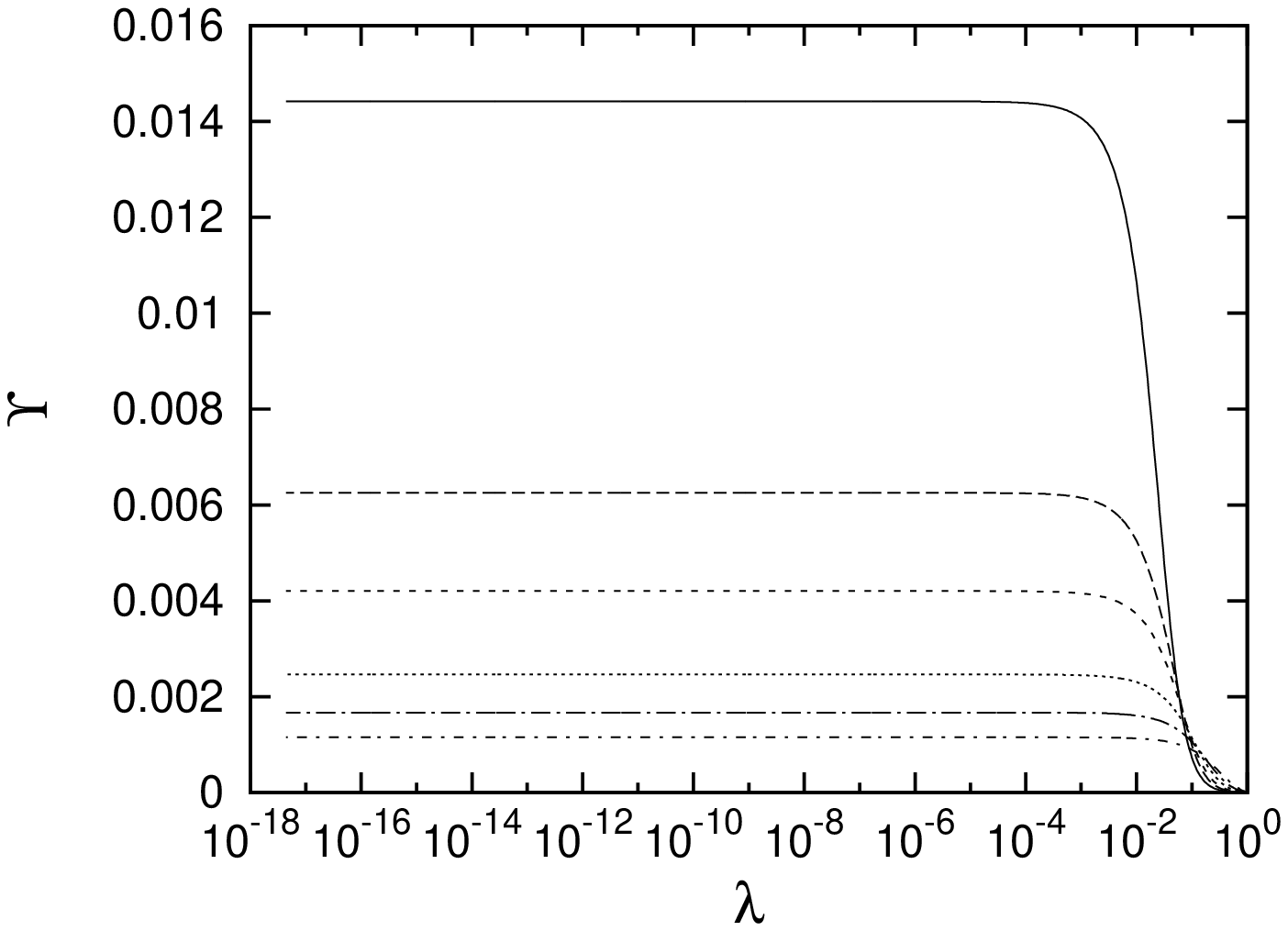,width=7cm,angle=-90}
\caption{\label{fig:nwfy}
The scale dependence  of  $\Upsilon$ for $\Omega^2_B=0.001,~g_B=0.01$,
and $\yb=\Yb=0$.  The order of the  curves from the top correspond to the various
 values  $N=1,10,20,50,100,200$, 
respectively.
} 
\end{center}
\end{figure}

The flows in the AWF have common features that can be recognized from 
Figs. \ref{fig:appr}-\ref{fig:nwfy}. Since the couplings $\yb$ and $\Yb$ scale quite 
similarly, we have plotted their combination $\Upsilon$ into which they merge in the case 
of the one-dimensional oscillator. After an UV scaling region
extending roughly over three orders of magnitude of the scale parameter $\lambda$, the 
flow saturates in the IR region for any of the dimensionful
couplings, so that all couplings  are IR relevant.  Namely, they tend to nonvanishing constant
values in the IR limit and only a single phase is  detected. The latter corresponds to the general expectation 
because there is no room for spontaneous symmetry  breaking in the $0+1$ dimensional quantum system.

\begin{figure}[ht] 
\begin{center} 
\epsfig{file=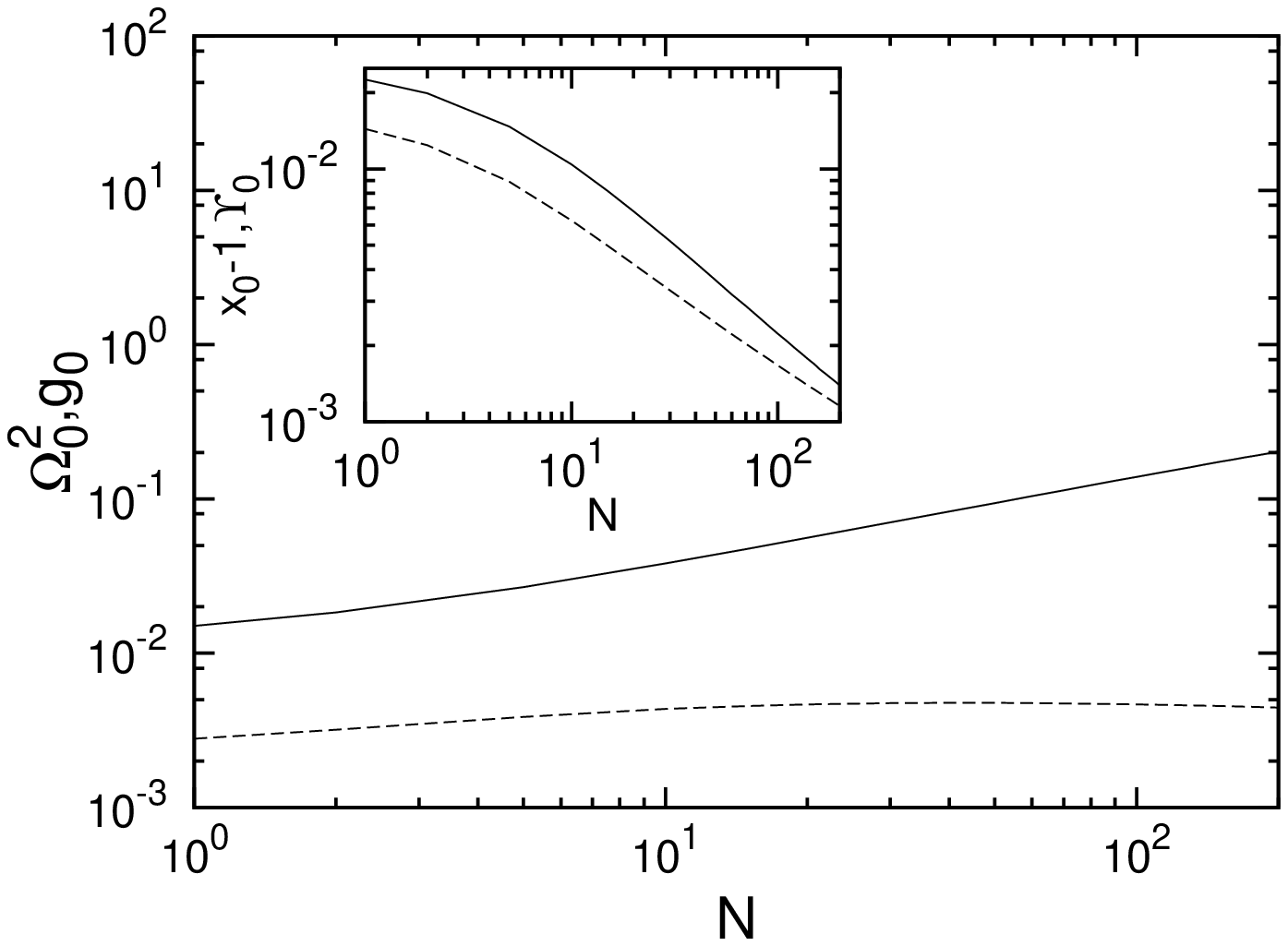,width=7cm,angle=-90}
\caption{\label{fig:nwf}
The $N$-dependence of the IR values  $\Omega^2_0$ (full line) and $g_0$ (dashed line)
for $\Omega_B^2=0.001$, $g_B=0.01$, and $\yb=\Yb=0$. The IR values of the wavefunction 
renormalization parameters
 $x_0$ (full line) and $\Upsilon_0$ (dashed line) are shown vs. $N$ in the inset. 
} 
\end{center}
\end{figure}

It has been found that the flows for any of the couplings do not alter qualitatively with increasing $N$.
That is illustrated in Figs. \ref{fig:nwfx} and \ref{fig:nwfy} for the parameters of the wavefunction 
renormalization.  The $N$-dependence of the various couplings in the IR limit is depicted in 
\fig{fig:nwf} for the various parameters. In this respect the IMA turned out to be misleading.
According to the IMA  the quantum fluctuations in the IR limit result in additive contributions
to the bare values of the couplings, which are linear in the dimension $N$, but the AWF gives
completely different behavior. We see  in \fig{fig:nwf} that the parameter 
$\Omega^2_0$ shows up a powerlike dependence on $N$ for  sufficiently large
dimensions,  $\Omega^2_0\propto N^{b_\Omega}$ with $b_\Omega\approx 0.58$.
In field-theoric terms the oscillator gets more `massive' with increasing dimension $N$.
The coupling $g_0$ depends rather weakly on  $N$ having a maximum at around 
$N\approx 30$. The slow decrease for large $N$ values may be the consequence of the
truncations used. The neglection of higher-order monomials in the gradient 
expansion can cause a failure when one or both of the dimension $N$ 
and the bare quartic coupling $g_B$  are too large.
It is easy to notice  from the explicit form of the RG evolution equations in the IMA that the
loop-expansion goes effectively with $Ng_B\hbar$ for large $N$, so that the higher-order
terms of the local potential can grow up to the same order of magnitude as the quadratic 
and quartic terms are when the flow 
reaches the border of the UV scaling region and that may result in their saturation in the 
IR scaling region at values of the same order of magnitude, too. 
In Refs. \cite{Aoki2002,Kapo2000} the authors take with the higher-order monoms of the
local potential, as well. They have found that the renormalized coupling $g_0$ increases 
strictly monotonically with increasing bare parameter $g_B$  for the one-dimensional 
oscillator. Similarly,  $g_0$ should increase with the dimension $N$ of the oscillator 
also strictly  monotonically, because the the loop-expansion goes effectively with $Ng_B$ for large $N$. 
As one can see in the inset of \fig{fig:nwf}, the deviation $x_0-1$ of the field-independent
wavefunction renormalization parameter $x_0$ from its bare value as well as
the coupling $\Upsilon_0$ of the field-dependent wavefunction renormalization go to
zero according to the power laws, $x_0-1\sim N^{-b_x}$ and $\Upsilon_0\sim 
N^{-b_\Upsilon}$ with $b_x\approx 0.55$ and $b_\Upsilon\approx 0.45$, respectively for increasing $N$.
Therefore the wavefunction renormalization effects seem to be washed out in the limit $N\to\infty$
because of the increment of the mass gap.

\begin{figure}[ht] 
\begin{center} 
\epsfig{file=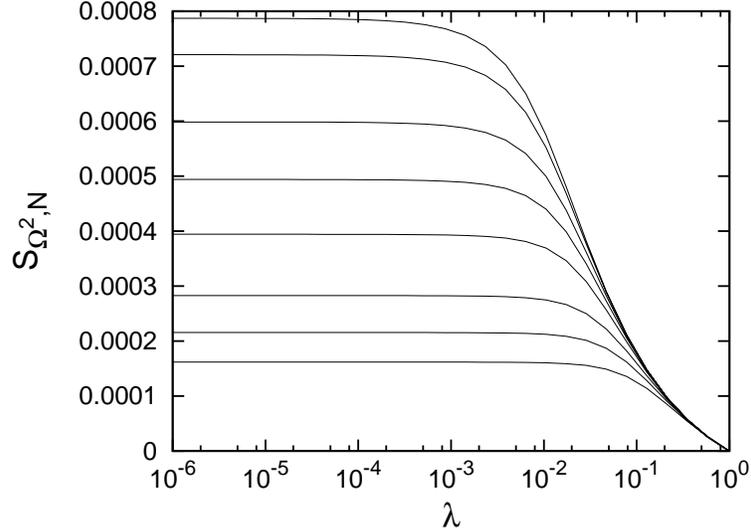,width=7cm,angle=-90}
\caption{\label{fig:sens1}
The flow of the sensitivity matrix element $S_{\Omega^2,N}$ for $\Omega^2_B=0.001$,
$g_B=0.00001$, and $\yb_B=\Yb_B=0$. The curves from above correspond to 
$N=1,2,5,10,20,50,100,200$, respectively.
}
\end{center}
\end{figure}

\begin{figure}[ht] 
\begin{center} 
\epsfig{file=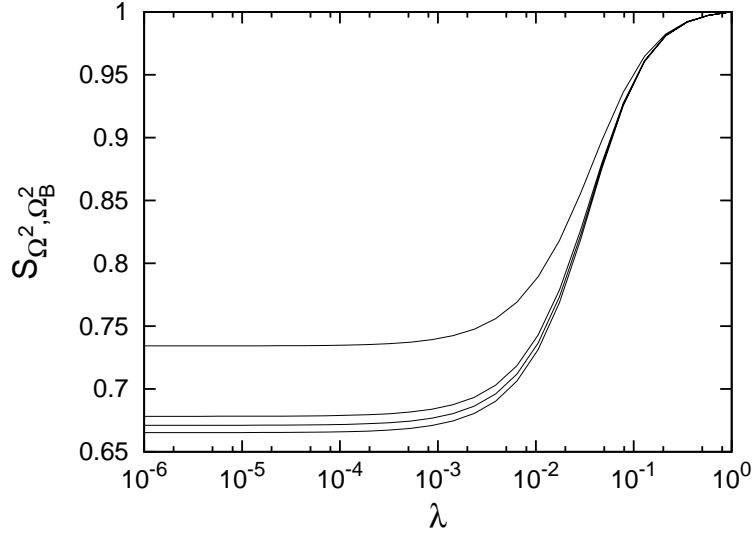,width=7cm,angle=-90}
\caption{\label{fig:sens2}
The flow of the sensitivity matrix element $S_{\Omega^2,\Omega_B^2}$ for $N=50$ and
$g_B=0.00001$. The curves from below correspond to $\Omega_B^2=-0.001,~-0.0001,~0.001,~0.01,$ respectively.
}
\end{center}
\end{figure}

\begin{figure}[ht] 
\begin{center} 
\epsfig{file=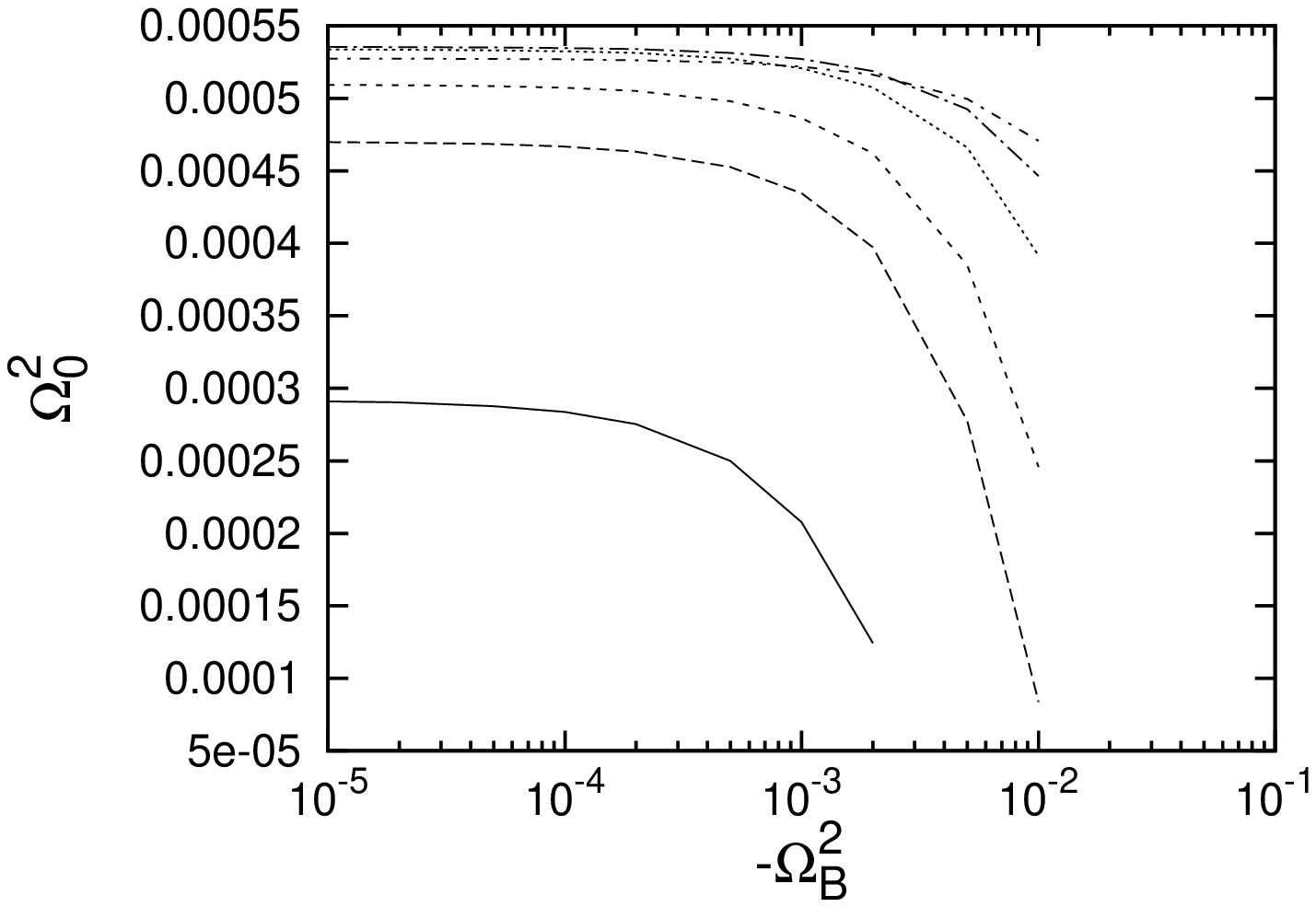,width=7cm,angle=-90}
\caption{\label{fig:dwn}
The $N$-dependence of the IR values $\Omega_0^2$ vs. the bare parameter
 $\Omega_B^2$.
The curves from below  correspond to $N=1,10,20,50,100,200$, respectively.
} 
\end{center}
\end{figure}

The dependence of the IR parameters on the bare ones can be characterized by the
sensitivity matrix, which is defined by
\be
S_{s_1,s_2} (r)= \frac{\partial s_1}{\partial s_2},
\ee
where $s_1=\Omega^2,~g,~ x,~\Upsilon$ are the running couplings and $s_2=
\Omega^2_B,~g_B,~ x_B,~\Upsilon_B,$ and $N$ are the bare parameters and the 
dimension, respectively. The elements of the sensitivity matrix integrate the infinitesimal
changes of the infinitesimal blocking steps during the RG flow when the ratio 
$r=\lambda/\lambda_B$ of the moving scale to the UV scale glides from $r=1$ to the IR
limit $r=0$.  Therefore, their 
values in the IR limit,  $S_{s_1,s_2}(0)$ as characteristics of the global RG flow may show
up singular dependences on the bare parameters, which may reveal themselves
in qualitatively different scale-dependences
$S_{s_1,s_2} (r)$ along the various RG trajectories started in various regions of the space
of the bare couplings \cite{Polo2004}. 
Since the model in our case has a single symmetric phase, no such singularities 
are expected to occur in the sensitivity matrix.
This  is demonstrated in Figs. \ref{fig:sens1} and \ref{fig:sens2}, where the matrix elements
$S_{\Omega^2,N}$ and $S_{\Omega^2,\Omega_B^2}$ are shown to vary 
monotonically with decreasing scale $\lambda$. The flow of these matrix elements do not
alter its character  with increasing $N$ (illustrated in \fig{fig:sens1}). The UV and the
IR scaling regions are again clearly distinguishable and the flow saturates in the IR region. 
The larger is the dimension $N$, the weaker is the sensitivity of the IR parameter 
$\Omega_0^2$ to the same small increment $\Delta N$ of the dimension.

Quite similarly, the character of the scale-dependences of $S_{\Omega^2,N}(r)$ and
$S_{\Omega^2,\Omega_B^2}(r)$ do not 
alter starting the RG flow with various bare values $\Omega_B^2$ (see \fig{fig:sens2}),
including also its negative values corresponding to a double-well bare potential.
This is another signal that the anharmonic oscillator exhibits only a single, symmetric phase.
As it was thoroughly discussed in \cite{Kapo2000,Aoki2002}  the large-amplitude, low-momentum
quantum fluctuations fill up the minima of the double-well potential and one ends up with 
a convex potential  and a unique ground state in the IR limit.
It was also discussed that  for sufficiently small bare quartic coupling the RG evolution
equations show up singularities and one has to take into account time-dependent 
saddle point  configurations for evaluating the path integral, that has been achieved by
turning to an instanton gas approximation.   
In \fig{fig:dwn} we illustrate that the RG flow started with negative parameter values
$\Omega_B^2<0$ ends up with some positive values $\Omega_0^2$ in the IR limit.
We note that in our case the evolution gets wrong for too large negative UV values of
$\Omega_B^2$ due to the strong truncation of the local potential in the ansatz \eq{ans}. 
As $N$ is increased we get larger and larger IR frequency values $\Omega_0^2$ according
to the power-law established previously.

Closing the discussion of the numerical results obtained in the  CSi RG  scheme, we can make
an estimate of the spatial extension of the ground-state wavefunction. As shown in the
Appendix \ref{a:csloc}, the variance of the coordinate operator $\u{q}$ can be estimated
by making use of the Thomas-Reiche-Kuhn sum rule \cite{Thom1925} and the dominance of the
first pole in the Lehmann-expansion of the single-particle propagator \cite{Aoki2002}.
With the help of the scaling of the renormalized coupling with increasing dimension $N$
established above, we get from \eq{qvar} that the variance scales as
\be\label{varscal}
   \langle 0 |\u{q}^2|0\rangle \approx 
\frac{Nx^{3/2} \hbar}{2\Omega} \sim N^{1-\hf b_\Omega}\sim N^{0.71} .
\ee
This means that with increasing dimension $N$ the quantum oscillator becomes more and more 
delocalized because of the increasing the number of degrees of freedom faster than that
of the `mass', ie. the frequency $\Omega_0$ of the oscillator.

\section{ISg RG for the one-dimensional oscillator}\label{sec:isrg}

\subsection{Evolution equations}

Further on we shall restrict ourselves to the study of the one-dimensional anharmonic oscillator,
since the scaling laws for the $N$-dimensional anharmonic oscillator do not show up
any significant qualitative modification in dependence on the number of dimensions $N$
according to our findings in the framework of the CSi RG scheme.
Now we  choose an IS RG scheme in that  the bare coupling $g_B$ of the
anharmonic quartic interaction term acts as control parameter and it is evolved from zero
towards a final positive value. For the sake of simplicity, below we shall suppress its index 
and denote the control parameter of the evolution simply by $g$.
Here we apply the ISg RG scheme to the so-called second  Legendre-transform, the Legendre-transform of the 
generating functional $W[J]$ of the connected Green-functions of the bilocal operator
$q_tq_{t'}$  defined via the path integral 
\bea
 e^{\ih W[J]}&=& 
\int \cD q e^{ \ih \int_t [\hf {\dot q}^2 - \hf (\omega_0^2-i\epsilon) q^2 
- \frac{g}{24} q^4]
    + \ih \hf \int_{t,t'}q_t J_{t,t'} q_{t'}}
\eea
with the symmetric external bi-local source $J_{t,t'}=J_{t',t}$ \cite{Polo2005}. We have introduced
the infinitesimal imaginary piece of the quadratic term of the potential by the
replacement $\omega_0^2\to \omega_0^2 -i\epsilon$ in order to make the path integral convergent. 
In the presence of the external source $J$ the propagator $G$ is given as
\be
W^{(1)}_{t,t'}=\fd{ W[J]}{\hf J_{t,t'}}= \la T( q_tq_{t'})\ra = G_{t,t'},~~~~
W^{(1)}_{\omega,\omega'}=\fd{W[J]}{\hf J_{-\omega,-\omega'}}  = G_{\omega,\omega'}.
\ee
The vanishing source $J=0$ corresponds to the ground state with the propagator
$\la T( q_tq_{t'})\ra_{J=0}=G_{\mr{gr}~t,t'}$. 
The second Legendre-transform,  the 2PI effective action $\Gamma[G]$  is defined as
\be
 \Gamma[G]=- W[J]+\hf \tr(G^\mr{tr}\cdot J),
\ee
and the relations
\bea
&&\fd{\Gamma[G]}{G_{t,t'}} = \hf J_{t,t'},~~~~
\fd{\Gamma[G]}{G_{\omega,\omega'}}= \hf  J_{-\omega,-\omega'}
\eea
hold. The symmetry of the source $J$ implies the symmetry $G_{t,t'}=G_{t',t}$ of 
the  propagator.  The second functional derivatives satisfy the identity
\bea
  I&=&  W^{(2)}: \Gamma^{(2)},~~~~ W^{(2)}=\Gamma^{(2)-1I}
=\Gamma^{(2)-1}:I
\eea
in shorthand notations with $A:B=\int_{t,t'} A_{(.,.),(t,t')}B_{(t,t'),(.,.)}
=\int_{\omega,\omega'}A_{(.,.),(\omega,\omega')}B_{(-\omega,-\omega'),(.,.)} $.
For later convenience we introduce the matrices $I$, $L$, and $\openone$,
\bea
I_{(t_3,t_4),(t_1,t_2)}&=&
 \hf (\delta_{t_1,t_3}\delta_{t_2,t_4}
+\delta_{t_1,t_4}\delta_{t_2,t_3}),~~~~I_{(\omega_3,\omega_4),(\omega_1,\omega_2)}=\hf
   (\delta_{\omega_1+\omega_3,0}\delta_{\omega_2+\omega_4,0}
   + \delta_{\omega_1+\omega_4,0}\delta_{\omega_2+\omega_3,0}),\nn
  L_{(t_3,t_4),(t_1,t_2)}&=&\delta_{t_1,t_2}\delta_{t_1,t_3}\delta_{t_1,t_4},~~~~
   L_{(\omega_3,\omega_4),(\omega_1,\omega_2)}
=\delta_{\omega_1+\omega_2+\omega_3+\omega_4,0},\nn
\openone_{(t_3,t_4),(t_1,t_2)}
&=&\delta_{t_1,t_3}\delta_{t_2,t_4},~~
\openone_{(\omega_3,\omega_4),(\omega_1,\omega_2)}=
\delta_{\omega_1+\omega_3,0}\delta_{\omega_2+\omega_4,0}.
\eea
The matrix $I$ acts over the two-particle (in field-theoric sense) subspace as the projector on the symmetrical
subspace (that of the two-particle states being symmetric under the exchange of the 
particles) and plays the role of the identity operator in that subspace and $   L:I=L$ holds, 
while $\openone$ represents the usual identity operator over the entire two-particle subspace. 
The inverse $A^{-1}$ of an arbitrary two-particle matrix $A$ satisfies $A^{-1}:A=
\openone$, while for the inverse $A^{-1I}=A^{-1}:I$ of the same matrix in the 
symmetrical subspace $A^{-1I}:A=I$ holds. Further on let us denote the trace of an arbitrary 
two-particle matrix by $\Tr A=\int_{t,t'}A_{(t,t'),(t,t')}=\int_{\omega,\omega'}
A_{(\omega,\omega'),(-\omega,-\omega')}$.

The bare coupling $g$  is turned on gradually   from zero to some finite
value in order to control the evolution of the 2PI effective action satisfying the
following RG evolution equation, 
\bea\label{gev}
\partial_g \Gamma[G]&=& -\partial_g W[J]=
  \frac{1}{24}  e^{-\ih W[J]} \int \cD q \int_t q_t^4 
 e^{ \ih \int_t ( \hf {\dot q}^2 - \hf (\omega_0^2-i\epsilon) q^2 - \frac{g}{24} q^4)
    + \ih \hf q\cdot J\cdot q}\nn
&=&
 \frac{1}{24}\biggl(\hi\biggr)^2 \biggl[ \ih\Tr (L:W^{(2)}) + \biggl(\ih\biggr)^2 W^{(1)}:L:W^{(1)}\biggr]
=
  \frac{1}{24} \hi
\biggl[ \Tr( L:\Gamma^{(2)-1I}) +\ih G:L:G\biggr].
\eea
Separating off the trivial term controlling the RG evolution, one can introduce the reduced 
2PI effective action $\Gab[G]$ via the relation
\bea\label{gab}
  \Gamma[G]= \Gab[G] + \frac{g}{24} G:L:G
\eea
and recast Eq. \eq{gev} in the form
\bea\label{gabev}
  \partial_{g} \Gab [G]&=&
 \frac{1}{24}\hi\Tr\biggl[ \biggl(
   \Gab^{(2)}[G] +\frac{g}{12}L \biggr)^{-1I}:L\biggr].
\eea

\subsection{Solution of the evolution equation}

\subsubsection{IMA}\label{dlegima}

In order to make a guess on the functional form of the reduced 2PI effective action $\Gab[G]$
for the linear quartic oscillator,
we evaluate it first in the tree-approximation, and solve the RG equation in the IMA.
The tree-level approximation corresponds to the vanishing value of
the control parameter, $g=0$, ie. the case of the linear harmonic oscillator, for which the
Gaussian path integral can be evaluated in a straightforward manner, yielding
\bea
  W^{\mr{tree}}[J] &=&
 \frac{i\hbar}{2} \tr \ln [   (\omega^2 -\omega_0^2+i\epsilon)\Ib + J]   
\eea
and
\bea
 \Gamma^{\mr{tree}}[G]
&=&
  \frac{i\hbar}{2} \tr \ln G  -\hf \tr  [ (\omega^2 -\omega_0^2+i\epsilon)\Ib\cdot G  ]
+{\rm{~const.~}}=\Gab^{\mr{tree}}[G]
 \eea 
up to an UV divergent additive constant (being independent of the bare parameters).
The necessary condition of the extremum of the 2PI effective action in the tree-approximation,
$\delta \Gab^{\mr{IMA}}[G]/\delta G=0$ provides the propagator in the ground state of the linear harmonic oscillator, 
\be
  G^{\mr{tree}}_{\mr{gr}}=  +i\hbar  [ (\omega^2 -\omega_0^2+i\epsilon)]^{-1} \Ib
\ee
as expected. The inverse of the noninteracting two-particle propagator is given via
\bea
   \cG^{{\mr{tree}}-1I}_{(-\omega_3,-\omega_4),(-\omega_1,\omega_2)}[G]&=&
  \Gab^{{\mr{tree}}(2)}_{(-\omega_1,-\omega_2),(-\omega_3,-\omega_4)}[G]
=
- \frac{i\hbar}{4}  [  G^{-1}_{-\omega_3,-\omega_1}  G^{-1}_{-\omega_2,-\omega_4}
+  G^{-1}_{-\omega_3,-\omega_2}  G^{-1}_{-\omega_1,-\omega_4} ].
\eea
Hence one finds
\be
  \cG^{\mr{tree}}_{(\omega_3,\omega_4),(\omega_1,\omega_2)}[G]=
    \ih ( G_{\omega_3,\omega_1} G_{\omega_4,\omega_2}+ 
 G_{\omega_4,\omega_1} G_{\omega_3,\omega_2} ) 
\ee
that we write  as $\cG^{\mr{tree}}=\ih (GG)^I$ introducing the shorthand notations
$(AB)^I_{(c,d),(a,b)}=
 \hf (A_{c,a}B_{d,b}+A_{d,a}B_{c,b})$, and for later use
$\alpha::\beta=\int_{a,\ldots,d} \alpha_{[\ldots],[(c,d),(a,b)]}
\beta_{[(-c,-d),(-a,-b)],[\ldots]}$ where Latin indices stand for frequency indices.
For later use we also evaluate the first and second functional derivatives 
of $ \cG^{\mr{tree}}[G]$,
\bea
 \fd{ \cG^{\mr{tree}}_{(\omega_3,\omega_4),(\omega_1,\omega_2)} [G]}{G_{-\omega,-\omega'} } 
&=&
 \ih \hf \biggl(  [
\delta_{\omega_3+\omega,0}\delta_{\omega_1+\omega',0}G_{\omega_4,\omega_2}
 +\delta_{\omega_4+\omega,0}\delta_{\omega_2+\omega',0} G_{\omega_3,\omega_1} 
+ (\omega_4\Leftrightarrow \omega_3) ]+(\omega\Leftrightarrow \omega')\biggr),
\eea
\bea
 \fdd{ \cG^{\mr{tree}}_{(\omega_3,\omega_4),(\omega_1,\omega_2)} [G]}{G_{-\omega,-\omega'} }{G_{-\omega'',-\omega'''} } 
&=&
\frac{i}{4\hbar}\biggl[\biggl(  [
\delta_{\omega_3+\omega,0}\delta_{\omega_1+\omega',0}
\delta_{\omega_4+\omega'',0}\delta_{\omega_2+\omega''',0} 
 +\delta_{\omega_4+\omega,0}\delta_{\omega_2+\omega',0} 
    \delta_{\omega_3+\omega'',0}\delta_{\omega_1+\omega''',0}
\nn
&&
+ (\omega_4\Leftrightarrow \omega_3) ]+(\omega\Leftrightarrow \omega')\biggr)
+ (\omega''\Leftrightarrow \omega''')\biggr].
\eea

In order to obtain the 2PI effective action in the IMA, we have to insert
the tree-level expression $\Gab^{\mr{tree}}[G]$ into the right-hand side of the
evolution equation \eq{gabev} and integrate it over the control parameter
from the initial value $g=0$ to some finite value $g$. Then we get
\bea\label{gabima}
   \Gab^{\rm{IMA}} [G]&=&\Gab^{\mr{tree}}[G]+ \frac{1}{24}\hi \int_0^{g} d\gb
 \Tr\biggl[ \biggl(
   \Gab^{{\mr{tree}}(2)}[G] +\frac{\gb}{12}L \biggr)^{-1I}:L\biggr]
=
\Gab^{\mr{tree}}[G]-\frac{i\hbar}{2}  \Tr \biggl[ I:\ln \biggl( \openone
 + \frac{g}{12} L:   \cG^{\mr{tree}} [G]\biggr)\biggr]\nn
&=&
\Gab^{\mr{tree}}[G]+\frac{i\hbar}{2}  \sum_{n=1}^\infty \frac{(-1)^n}{n}
\biggl(  \frac{g}{12}\biggr)^n 
\tr \biggl (L:   \cG^{\mr{tree}} [G] \biggr)^n\biggr].
\eea
The inverse propagator of the ground state in the IMA is given through the necessary
condition of the extremum of the 2PI effective action $\Gamma^{\mr{IMA}} [G]$, ie. 
$\delta\Gamma^{\rm{IMA}} [G]/\delta G=0$, that yields the equation for the propagator
of the ground state in the IMA, $G^{\mr{IMA}}_{\mr{gr}}$,
\bea
  G^{\mr{IMA}~-1}_{\mr{gr}} &=& G_{\mr{gr}}^{\mr{tree}-1}
 + \frac{ g}{12} \Tr\biggl[
 L:\fd{ \cG^{\mr{tree}} [G]}{G}:\biggl( \openone+ \frac{g}{12}
L:\cG^{\mr{tree}} [G]\biggr)^{-1}\biggr]_{G=G^{\mr{IMA}}_{\mr{gr}}}
+ \frac{ig}{6\hbar} (L:G^{\mr{IMA}}_{\mr{gr}}),
\eea
ie.
\bea\label{sdima}
 G^{\mr{IMA}~-1}_{\mr{gr}~\omega,\omega'} 
&=&
 G_{\mr{gr}~\omega,\omega'}^{\mr{tree}-1} +\biggl[\frac{ig}{2\hbar}
\int_{\omega_1,\omega_2}\delta_{\omega+\omega'+\omega_1+\omega_2,0}
G_{\omega_1,\omega_2}\nn
&&
  + \frac{g^2}{18\hbar^2} \int_{\omega_1,\omega_2,\ldots,\omega_6}
 \delta_{\omega+\omega_1+\omega_3+\omega_4,0}\delta_{\omega'+\omega_2
+\omega_5+\omega_6,0} G_{\omega_1,\omega_2}G_{\omega_3,\omega_5}G_{\omega_4,\omega_6}+\ord{g^3}
\biggr]_{G=G_{\mr{gr}}^{\mr{IMA}}}
\eea
when the terms up to the order $\ord{g^2}$ are only kept.
Multiplying both sides of this equation by $G^{\rm{IMA}}_{\mr{gr}}$ from the left
and $G_{\mr{gr}}^{\mr{tree}}$ from the right, one would obtain the Schwinger-Dyson equation in the IMA for the
propagator of the ground state, but for our later purposes
it is more convenient to retain the present form of the equation. 
The solution of Eq. \eq{sdima}, the propagator of the ground state can be used to determine
the energy of the ground state,
\be
  E_0= \lim_{T\to \infty}\frac{1}{T}  \Gamma^{\mr{IMA}}[G_{\mr{gr}}^{\mr{IMA}}],
\ee
and the 2PI  4-point vertex function,
\bea\label{ga2ima}
\lefteqn{
  \gamma^{[2]~\mr{IMA}}_{(\omega_3,\omega_4),(\omega_1,\omega_2)}
=
\cG^{\rm{IMA}-1I}_{(\omega_3,\omega_4),(\omega_1,\omega_2)}[G_{\mr{gr}}^{\mr{IMA}}]
-\cG^{\rm{tree}-1I}_{(\omega_3,\omega_4),(\omega_1,\omega_2)}[G^{\mr{IMA}}_{\mr{gr}}]
}\nn
&=&
\frac{g}{4}\delta_{\omega_1+\omega_2+\omega_3+\omega_4,0} -\frac{ig^2}{24\hbar}
\int_{\omega_1',\omega_2',\omega_3',\omega_4'}[\delta_{\omega_1+\omega_3+\omega_1'+\omega_3',0}
  \delta_{\omega_2+\omega_4+\omega_2'+\omega_4',0}+  (\omega_1\Leftrightarrow \omega_2) ]
G^{\rm{IMA}}_{\rm{gr}~\omega_1',\omega_2'}G^{\rm{IMA}}_{\rm{gr}~\omega_3',\omega_4'}+\ord{g^3}\nn
\eea
with $\cG^{\rm{IMA}-1I}[G]=\Gamma^{\rm{IMA}~(2)}[G]$.
\begin{figure}[thb]
\psfrag{LL}{\Large $L=$}
\psfrag{gL}{\Large $gL=$}
\psfrag{gg2}{\Large $\gamma^{[2]}=$}
\psfrag{gg4}{\Large $\gamma^{[4]}=$}
\psfrag{calG}{\Large ${\cal G}=$}
\psfrag{pGam}{\Large $\partial\Gamma=$}
\psfrag{GmGt}{\Large $\Gamma-\Gamma^{tree}=$}
\psfrag{eqq}{\Large $=$}
\psfrag{ordg4}{\Large $\ord{g^4}$}
\psfrag{ordg3}{\Large $\ord{g^3}$}
\includegraphics[width=10cm]{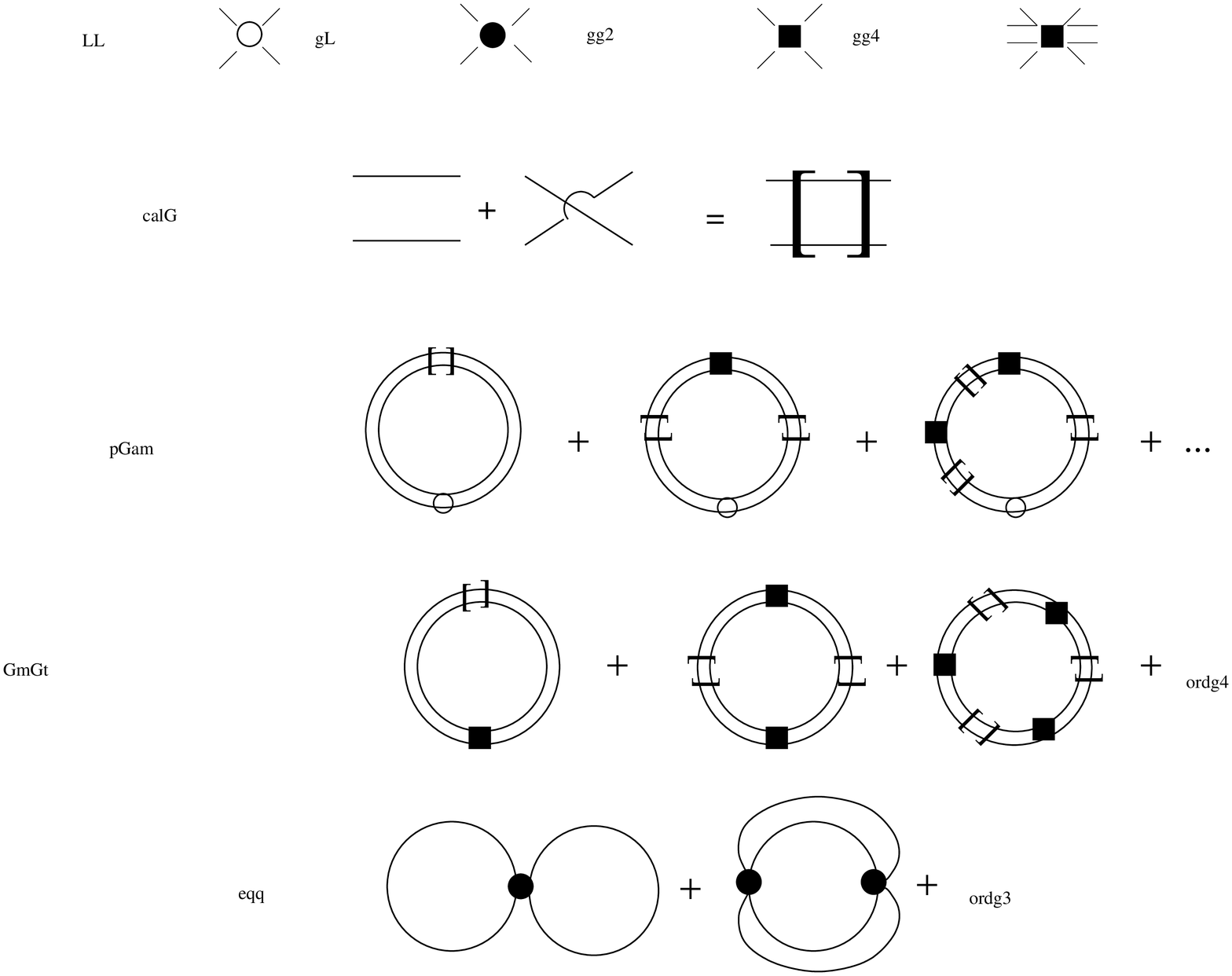}
\caption{Feynman diagrams for the RG evolution of the 2PI effective action in the IMA.
\label{fig:dleg}}
\end{figure}
\begin{figure}[thb]
\psfrag{ordg3}{\Large $\ord{g^3}$}
\psfrag{sig}{\Large $\Sigma=$}
\psfrag{gam2}{\Large $\gamma^{[2]}=$}
\includegraphics[width=10cm]{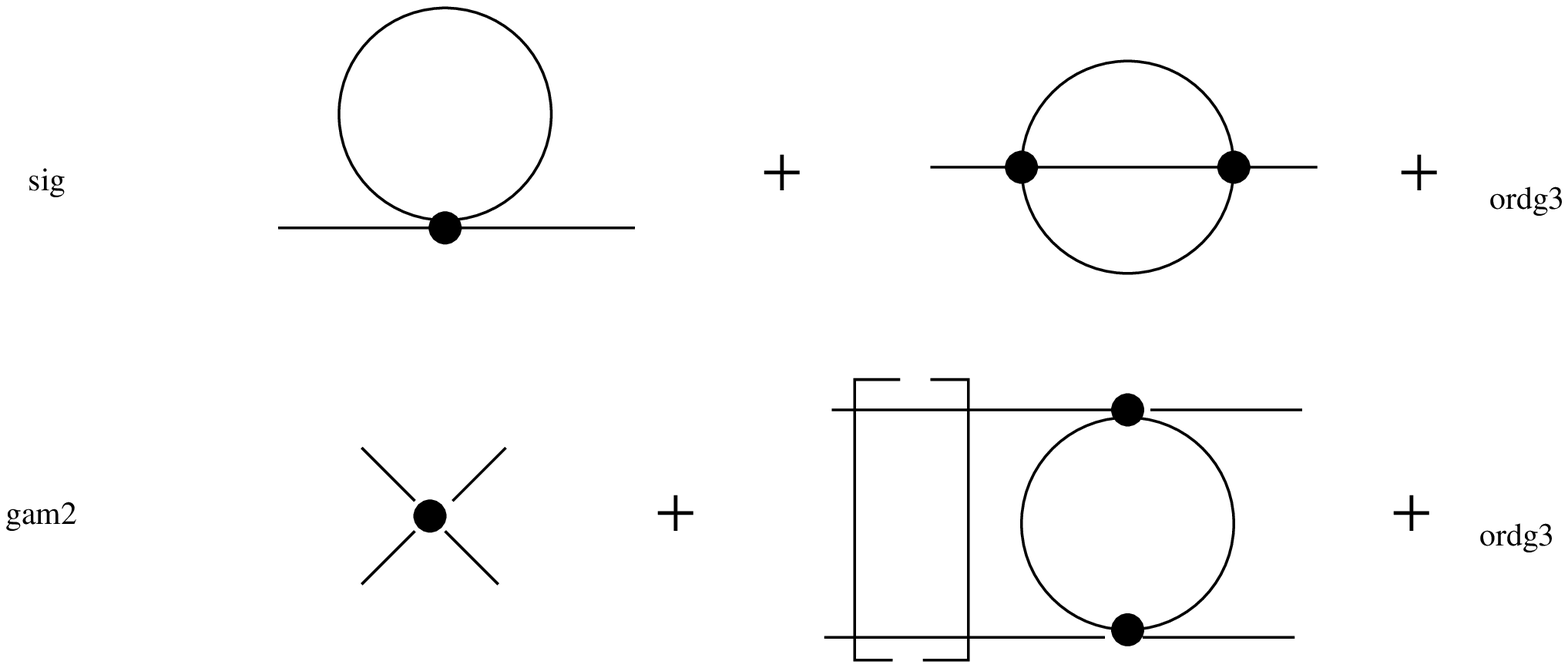}
\caption{Feynman diagrams for the proper self-energy insertion $\Sigma=G^{-1~\mr{IMA}}
-G^{-1~\mr{tree}}$ and the 2PI vertex function $\gamma^{[2]~\mr{IMA}}$.
\label{fig:sefveima}}
\end{figure}

The evolution equation \eq{gabev} allows for a simple diagrammatic picture, shown in Fig. \ref{fig:dleg}.
The single vertex  without the factor $g$ (empty circle) present in each of the diagrams 
for $\partial_g \Gab$ occurs due to the partial derivative with respect to $g$. The result of the IMA is the 
resummation of the ring diagrams shown for $\Gamma[G]-\Gamma^{\mr{tree}}[G]$ in the fourth and
fifth lines in Fig. \ref{fig:dleg}.  In  the fourth line we depicted the expansion of the 2PI effective
action in powers of the tree-level two-particle propagator $\cG^{\mr{tree}}$, each 
diagram consists of double-line sections for $\cG^{\mr{tree}}$'s  joined via the 
vertex functions $\gamma^{[2]~\mr{tree}}$ to a closed loop. As one can see from the
expansion  in powers of the propagator $G$  shown in the fifth line, the 2PI vertex function
is taken at the tree-level, $\gamma^{[2]~\mr{tree}}=gL/12$.
The proper self-energy insertion $\Sigma^{\mr{IMA}}= G^{-1~\mr{IMA}}-G^{-1~\mr{tree}}$ 
and the 2PI 4-point vertex function $\gamma^{[2]~\mr{IMA}}$ in the IMA are depicted in 
Fig. \ref{fig:sefveima}.

\subsubsection{Expansion in the 2-particle channel (E2PC)}\label{dleg2lch}

The functional dependence of the reduced 2PI effective action $\Gab^{\rm{IMA}}[G]$ on
the  tree-level 2-particle propagator $\cG^{\mr{tree}}$ in the IMA brings one to the
idea to look for a better approximation of the effective action as an expansion in powers of
the tree-level 2-particle propagator $\cG^{\mr{tree}}=2\ih (GG)^I$ making the Ansatz
\bea\label{efexp}
  \Gab^{\mr{E2PC}}[G]
&=&\Gamma^{\mr{tree}}[G] + \hf \gab^{[2]}::(GG)^I 
+\frac{1}{24} (GG)^I::\gamma^{[4]}::(GG)^I+\ldots,
\eea
where the symmetry properties
\bea
 &&\gab^{[2]}_{(a,b),(c,d)}=\gab^{[2]}_{(b,a),(c,d)}=\gab^{[2]}_{(c,d),(a,b)},\nn
&& \gamma^{[4]}_{[(a,b),(c,d)],[(a',b'),(c',d')]}=
   \gamma^{[4]}_{[(b,a),(c,d)],[(a',b'),(c',d')]}=
\gamma^{[4]}_{[(c,d),(a,b)],[(a',b'),(c',d')]}
=\gamma^{[4]}_{[(a',b'),(c',d')],[(a,b),(c,d)]}
\eea 
are required (Latin indices stand for frequency indices) for the 2PI  vertex functions, 
$\gab^{[2]}$ and $\gamma^{[4]}$.
We shall solve the evolution equation  \eq{gabev} inserting the Ansatz \eq{efexp}
into it, expanding its right-hand side in Neumann-series,
and keeping the terms up to the quartic ones in $G$, ie. those of  the order
$\ord{[(GG)^I]^2}$ in its both sides. The approximation with the E2PC  goes beyond
the IMA when   the inverse of the 2-particle propagator, $\Gab^{(2)}$ has been replaced 
by $\Gamma^{\mr{tree}(2)} = \cG^{\mr{tree}-1}[G]$
in the right-hand side of the RG equation \eq{gabev}. Now the Ansatz \eq{efexp}
in powers of $\cG^{\mr{tree}}[G]$ implies 
\be
 \Gab^{\mr{E2PC}~(2)}[G] =\cG^{\mr{tree}-1}[G]+ \gab^{[2]}+\ord{(GG)^I}.
\ee
Having expanded the inverse matrix, ie. 2-particle propagator in the right-hand side of \eq{gabev} in 
Neumann-series as
\be\label{inv2lch}
 \cG^{\mr{E2PC}}[G]=\Gamma^{\mr{E2PC}(2)~-1}[G]=
 \biggl( \Gab^{\mr{E2PC}~(2)}[G]+\frac{g}{12}L \biggr)^{-1}=
  \cG^{\mr{tree}}[G] - \cG^{\mr{tree}}[G]: \gamma^{[2]} :\cG^{\mr{tree}}[G]+ \ord{[(GG)^I]^3}
\ee
with the 2PI 4-point vertex function,
\be\label{ga2lch}
\gamma^{[2]}= \Gamma^{(2)}[G]-\cG^{\mr{tree}-1}[G]
= \gab^{[2]}+\frac{g}{12}L ,
\ee
we obtained the following coupled set of RG evolution equations for the 2PI 4-point and
8-point vertex functions,
\bea
  \partial_g \gab^{[2]}_{(\omega_1,\omega_2),(\omega_3,\omega_4)}
&=&\frac{1}{6}L_{(\omega_3,\omega_4),(\omega_1,\omega_2)},
\nn
 \partial_g\gamma^{[4]}_{[(\omega_1,\omega_2),(\omega_3,\omega_4)],[(\omega_1',\omega_2'),(\omega_3',\omega_4')]}&=&
  -\frac{i}{\hbar}\biggl[
 \biggl((I:\gamma^{[2]}:I)_{(\omega_3,\omega_4),(\omega_3',\omega_4')}L_{(\omega_1,\omega_2),(\omega_1',\omega_2')}\nn
&&
                               + L_{(\omega_3,\omega_4),(\omega_3',\omega_4')}(I:\gamma^{[2]}:I)_{(\omega_1,\omega_2),(\omega_1',\omega_2')}
\biggr)+ (\omega_1,\omega_2)\Leftrightarrow (\omega_3,\omega_4)\biggr] .
\eea
Those can be integrated straightforwardly
for the initial conditions $\gab^{[2]}_{g=0}=0,~\gamma^{[4]}_{g=0}=0$, and one gets
\bea\label{ga24lch}
\gab^{[2]}_{(\omega_1,\omega_2),(\omega_3,\omega_4)}&=& \frac{g}{6}L_{(\omega_1,\omega_2),(\omega_3,\omega_4)},\nn
  \gamma^{[4]}_{[(\omega_1,\omega_2),(\omega_3,\omega_4)],[(\omega_1',\omega_2'),(\omega_3',\omega_4')]}&=&
-\frac{i g^2}{4\hbar}[\delta_{\omega_1+\omega_2+\omega_1'+\omega_2',0}
\delta_{\omega_3+\omega_4+\omega_3'+\omega_4',0}
+ \delta_{\omega_3+\omega_4+\omega_3'+\omega_4',0}\delta_{\omega_1+\omega_2+\omega_3'+\omega_4',0}] 
\eea
and hence $\gamma^{[2]}=\frac{g}{4}L$. Since we have kept only the contribution of $\ord{g}$
to  the reduced 2PI 4-point vertex function $\gab^{[2]}$,
the expansion \eq{efexp} of the functional $\Gab^{\mr{E2PC}}[G]$
in  powers of the free 2-particle propagator $\cG^{\mr{tree}}\sim (GG)^I$ corresponds
to its expansion in powers of $g$. The 2PI effective action takes now the form
\bea\label{efac2ch}
  \Gamma^{\mr{E2PC}}[G]&=&
\Gamma^{\mr{tree}}[G]
+ \frac{g}{8} \int_{\omega_1,\ldots,\omega_4}\delta_{\omega_1+\omega_2+\omega_3+\omega_4,0}G_{\omega_1,\omega_2}G_{\omega_3,\omega_4}\nn
&&
- \frac{ig^2}{48\hbar}\int_{\omega_1,\ldots,\omega_4,\omega_1'\ldots,\omega_4'}
 \delta_{\omega_1+\omega_2+\omega_1'+\omega_2',0}\delta_{\omega_3+\omega_4
+\omega_3'+\omega_4',0}G_{\omega_1,\omega_3}G_{\omega_2,\omega_4}G_{\omega_1',\omega_3'}G_{\omega_2',\omega_4'} +\ord{G^6}.
\eea
In this approximation based on the E2PC there occurs an additional contribution of the
order $\ord{g^2}$ in the 2PI effective action which was not included  in the IMA:
in the right-hand side of Eq.  \eq{efac2ch} there occurs an additional factor 3 in the term 
of the order $\ord{g^2}$ as compared  to the corresponding term in the IMA. 
This is because of the improvement of the 2-particle propagator,
\bea
  \cG^{\mr{E2PC}}[G]&=& 
  \cG^{\mr{tree}} [G]+ \frac{g}{4}  \cG^{\mr{tree}}[G]:L:\cG^{\mr{tree}}[G]
+\ord{g^2}
\eea 
(c.f. Eq. \eq{inv2lch}) as compared to the free one used in the IMA. Putting it in another way,
in the approximation base on the E2PC we use the 2PI 4-point vertex function
$\gamma^{[2]}=\gab^{[2]}+\gamma^{[2]~\mr{tree}}=(g/4)L$ instead of $\gamma^{[2]~\mr{tree}}=(g/12)L$
used in the IMA. The diagrammatic form   of RG the evolution equation remains the same as that in the IMA
(see Fig. \ref{fig:dleg}) except of replacing the vertices $\gamma^{[2]~\mr{tree}}$ by $\gamma^{[2]}$
of the present approximation.

The necessary condition of the extremum of the 2PI effective action,
$\delta \Gamma^{\mr{E2PC}}[G]/\delta G=0$ provides again the
Schwinger-Dyson equation for the propagator of the ground state,
\bea\label{prgr2lch}
  G^{\mr{E2PC}~-1}_{\mr{gr}~\omega_1,\omega_2}&=&
G_{\mr{gr}~\omega_1,\omega_2}^{\mr{tree}-1}  
 +\biggl[  \frac{ig}{2\hbar}  \int_{\omega_1',\omega_2'}\delta_{\omega_1+\omega_2
+\omega_1'+\omega_2',0}G_{\omega_1',\omega_2'}
\nn
&&+\frac{g^2}{6\hbar^2}\int_{\omega_1',\ldots,\omega_4',\omega_3,\omega_4}
\delta_{\omega_1+\omega_3+\omega_1'+\omega_2',0}\delta_{\omega_2+\omega_4
+\omega_3'+\omega_4',0}G_{\omega_3,\omega_4}G_{\omega_1',\omega_3'}G_{\omega_2',\omega_4'}\biggr]_{G= G^{\mr{E2PC}}_{\mr{gr}}}
\eea
and a straightforward but lengthy calculation yields 
\bea\label{g2lch2lch}
\lefteqn{
\gamma^{[2]~\mr{E2PC}}_{(\omega_3,\omega_4),(\omega_1,\omega_2)} =
\cG^{\mr{E2PC}~-1I}_{(\omega_3,\omega_4),(\omega_1,\omega_2)}[G_{\mr{gr}}^{\mr{E2PC}}]
 -\cG^{\mr{tree}~-1I}_{(\omega_3,\omega_4),(\omega_1,\omega_2)} [G_{\mr{gr}}^{\mr{E2PC}}]
}
\nn
&=&
 \frac{g}{4}\delta_{\omega_1+\omega_2+\omega_3+\omega_4,0}
 -\frac{ig^2}{8\hbar}\int_{\omega_1',\omega_2',\omega_3',\omega_4'}
[\delta_{\omega_1+\omega_3+\omega_1'+\omega_3',0}\delta_{\omega_2+\omega_4
+\omega_2'+\omega_4',0}+(\omega_1\Leftrightarrow \omega_2)]
[ G_{\omega_1',\omega_2'}G_{\omega_3',\omega_4'}]_{G=G_{\mr{gr}}^{\mr{E2PC}} }
\eea
for the value of the 2PI 4-point  vertex in the ground state. The $\ord{g^2}$ term of the
 propagator as well as 
that of the 2PI vertex function  exhibit the additional factor 3 as compared to
 the corresponding terms in the IMA.

\subsection{Determination of the renormalized couplings}

We shall obtain the renormalized quantities in the IS RG scheme in the IMA and in the truncated E2PC,
as described previously in Sections \ref{dlegima} and \ref{dleg2lch}, respectively, in both
cases keeping the terms up to the second order $\ord{g^2}$ of the bare coupling $g$ that
has been used as the control parameter. We shall make use of the diagonal form of the 
propagator, $G_{\omega,\omega'}=G_\omega \delta_{\omega,\omega'}$.
First, we determine the propagator $G_{\mr{gr}~\omega}$ of the ground state by solving
the Schwinger-Dyson equation with the precision $\ord{g^2}$, and afterwards the 
energy of the ground state via
\be
  E_0= \lim_{T\to \infty}\frac{1}{T}  \Gamma[G_{\mr{gr}}],
\ee
and the renormalized coupling $g_0$ from the 2PI 4-point vertex function taken at the
symmetric point $\omega_i=0$ $(i=1,2,3,4)$ via
\be\label{gr}
   g_0= \lim_{T\to \infty,\omega_i\to  0~}\frac{4}{T}
\gamma^{[2]}_{(\omega_3,\omega_4),(\omega_1,\omega_2)}.
\ee
Finally, we compare these results with those obtained at the same order of the perturbation expansion
in the CS RG scheme given in Appendix \ref{a:cspe},  in the WH RG scheme given in Appendix \ref{whpe},
as well as by the results of the well-known Rayleigh-Schr\"odinger perturbation
expansion in quantum mechanics. 

First, let us present the results obtained in the framework of the IS RG in the IMA. 
For diagonal propagator the Schwinger-Dyson equation \eq{sdima} can be
 rewritten as
\bea\label{sdima2}
G^{\mr{IMA}~-1}_{\mr{gr}~\omega}&=&
   G^{\mr{tree}~-1}_{\mr{gr}~\omega} 
+ \biggl[ \frac{ig}{2\hbar} \int_{\omega_1}G_{\omega_1}
 + \frac{g^2}{18\hbar^2}\int_{\omega_1,\omega_2,\omega_3}
\delta_{\omega+\omega_1+\omega_2+\omega_3,0}
G_{\omega_1}G_{\omega_2}G_{\omega_3} 
 +\ord{g^3} \biggr]_{G=G^{\mr{IMA}}_{\mr{gr}}}.
\eea
Let us insert into Eq. \eq{sdima2}  the perturbation expansion
\bea
  G^{-1}&=& G_0^{-1}+ gA+ g^2 B+  \ord{g^3},~~~~
G=G_0-gG_0\cdot A\cdot G_0-g^2G_0\cdot C\cdot G_0+  \ord{g^3}
\eea
where $A$ and $B$ are symmetric matrices to be determined, $C= B-A\cdot G_0\cdot A$
and we wrote $G_0=G^{\mr{tree}}_{\mr{gr}}$ for the sake of simplicity.
Comparing the terms of identical powers of the coupling $g$ in both sides of Eq. \eq{sdima2}, one finds 
$ A_{\omega,\omega'}=A_\omega\delta_{\omega+\omega',0}$ and
$B_{\omega,\omega'}=B_\omega\delta_{\omega+\omega',0}$ with
\bea
 && A_\omega=\frac{i}{2\hbar }
  \int_{\omega_1}  G_{0~\omega_1},~~~~
B_\omega=
\frac{1}{4\hbar^2 }
  \int_{\omega_1}G_{0~\omega_1}^2  \int_{\omega_2}  G_{0~\omega_2}
+ \frac{1}{18\hbar^2 }   \int_{\omega_1,\omega_2}
 G_{0~\omega_1+\omega_2-\omega}  G_{0~\omega_1}G_{0~\omega_2} .
\eea
The loop integrals can be taken explicitly making use of the residuum theorem,
\bea
 && \int_\omega G_{0~\omega}= \frac{\hbar}{2\omega_0},
~~~~
\int_\omega G_{0~\omega}^2 =\frac{-i\hbar^2}{4\omega_0^3},
~~~~
 \int_{\omega_1,\omega_2}
 G_{0~\omega_1+\omega_2-\omega}  G_{0~\omega_1}G_{0~\omega_2}
= \frac{3i\hbar^3}{4\omega_0^2}\frac{1}{\omega^2-9\omega_0^2+18i\epsilon}.
\eea
Expanding the last loop integral in powers of $\omega^2-\omega_0^2$ and keeping the 
terms up to the order $\ord{ \omega^2-\omega_0^2}$, one finds for the inverse of the propagator of the ground state
\bea\label{prgrima}
   G^{\mr{IMA}~-1}_{\mr{gr}~\omega}
&=&
 -\frac{i}{\hbar}( 
x_0\omega^2 -\Omega_0^2 +i\epsilon)
\eea
with
\bea
  && \Omega_0^2=\omega_0^2\biggl[1 +4\xi
-\frac{55}{6}\xi^2\biggr]+\ord{g^3},
~~~~
x_0= 1+\frac{1}{3}\xi^2+\ord{g^3}
\eea
in terms of the dimensionless parameter $\xi= \frac{g\hbar}{16\omega_0^3}$, introduced previously.
For diagonal propagator $G$ the 2PI effective action given by Eqs. \eq{gab} 
and  \eq{gabima}  and the 2PI 4-point  vertex function \eq{ga2ima} reduce to 
\bea\label{gima}
\Gamma^{\mr{IMA}}[G]&=&
\Gab^{\mr{tree}}[G] + \frac{g}{8}T \biggl(\int_\omega G_\omega\biggr)^2
-\frac{ig^2}{144\hbar}T\int_{\omega_1,\omega_2,\omega_3}G_{\omega_1 }
G_{\omega_2}G_{\omega_3}G_{\omega_1+\omega_2+\omega_3}+\ord{G^6}
\eea
and
\bea\label{g2liima}
  \gamma^{[2]~IMA}_{(\omega_3,\omega_4),(\omega_1,\omega_2)}
&=& \biggl[\frac{g}{4} - \frac{ig^2}{24\hbar} \int_{\omega} G_\omega
( G_{-\omega_1-\omega_3-\omega}+ G_{-\omega_2-\omega_3-\omega})
\biggr]\delta_{\omega_1+\omega_2+\omega_3+\omega_4,0},
\eea
respectively. Inserting $G=G^{\rm{IMA}}_{\mr{gr}}$ into the expression \eq{gima}
one finds for the energy of the ground state
\bea
E_0
&=&
\hf \hbar \omega_0 \biggl[ 1 +\xi
  - \frac{55}{9}\xi^2  +\ord{g^3}\biggr].
\eea
According to Eqs. \eq{gr} and \eq{g2liima}, one obtains for the renormalized quartic coupling,
\bea
  g_0&=& g- \frac{ig^2}{3\hbar }\int_\omega G_{0~\omega}^2
+\ord{g^3} = 
   g\biggl[1- \frac{4}{3}\xi +\ord{g^2}\biggr] .
\eea

At second, we turn to the approximation based on the E2PC.
Making use of  Eqs. \eq{efac2ch}, \eq{prgr2lch}, and \eq{g2lch2lch}, one can proceed like
in the case of the IMA and obtain  results with slightly different factors for the second-order terms,
\bea
 && \Omega_0^2=
\omega_0^2\biggl[1+ 4\xi
-\frac{71}{8}\xi^2 +\ord{\xi^3}\biggr]
,
~~~~
x_0= 1+ \frac{1}{8}\xi^2+\ord{\xi^3}
,
\eea
\bea
E_0 
&=&
  \hf\hbar\omega_0 \biggl[ 1+\xi
  -\frac{16}{3}\xi^2 +\ord{\xi^3}\biggr],
\eea
\bea
&& g_0
= g\biggl( 1- 4\xi+\ord{\xi^2}\biggr).
\eea

Finally, in Table \ref{tab:comp} we compare the results obtained in the various
approximation schemes. The various observables are given keeping the terms up to the
order $\ord{g_B^2}$ of the bare coupling. We compare here the renormalized values, ie. 
the IR limits of them obtained in the  CSi RG scheme  in second order of the perturbation
expansion as given in Appendix \ref{a:cspe}, the renormalized values obtained in the ISg RG
scheme in IMA and in the approximation based on the E2PC when the evolution has been ended 
up and the control parameter reached the same bare value $g_B$ that used in the CSi RG
scheme. We also present the results obtained in the framework of the WH
RG scheme as outlined in  Appendix \ref{whpe}, 
and the  results of the Rayleigh-Schr\"odinger perturbation expansion
(RSPE) \cite{Land1965}. In the RSPE one finds for the energy of the $n$-th energy niveau
$(n=0,1,2,\ldots)$ of the one-dimensional anharmonic oscillator with the bare Hamiltonian
$H=\hf p^2+ \hf \omega_0^2x^2+\frac{g_B}{24}x^4$,
\be
E_n=\hbar\omega_0\biggl[ n+\hf+e_n^{(1)}+e_n^{(2)}+\ord{\xi^3} \biggr]
\ee
with
\bea
 e_n^{(1)}&=&(1+2n+2n^2)\xi,\nn
 e_n^{(2)}&=& -\frac{1}{18} \xi^2\sum_{n'\not= n}\frac{1}{n'-n}[
 4(2n-1)^2 n(n-1)\delta_{n',n-2} + n(n-1)(n-2)(n-3)\delta_{n',n-4}\nn
&&
+ 4(2n+3)^2(n+1)(n+2)\delta_{n',n+2}+ (n+1)(n+2)(n+3)(n+4)\delta_{n',n+4}].
\eea
This yields for the first two energy levels
\bea
 E_0&=& \hf \hbar \omega_0\biggl( 1+\xi -\frac{7}{3}\xi^2 +\ord{\xi^3} \biggr),\nn
 E_1&=& \hf \hbar \omega_0\biggl( 3 +5\xi - \frac{55}{3}\xi^2+\ord{\xi^3}\biggr)
\eea
and
\be
 (\hbar \Omega_0)^2= (E_1-E_0)^2=
  (\hbar\omega_0)^2 (1 +4\xi -12\xi^2).
\ee
The same expression for the energy $E_0$ of the ground-state has been found in Ref. \cite{Haym1991}
by path-integral method. There are no asymptotically free states of the particle
moving in the quartic anharmonic oscillator potential, therefore the definition of scattering 
amplitudes and through those a renormalized quartic coupling is problematic in the RSPE
approach.

\begin{table}[thb]

\begin{center}
\begin{tabular}{|c|c|c|c|c|c|}
\hline 
     &  & & & & \cr
Observable                                                                           & ISg RG IMA                                          & 
    ISg RG     E2PC                                   & CSi    RG                               & WH RG& RSPE \cr
       &  & & & & \cr
\hline   
   &  & & & & \cr
  $ E_0 -\hf \hbar \omega_0-\xi$                                       &  $ -\frac{55}{9} \xi^2$     &     
    $-\frac{16}{3}\xi^2$        & $ -\frac{71}{30}\xi^2$     & $ -18\xi^2 $      & $-\frac{7}{3}\xi^2$   \cr 
   &  & & & & \cr
\hline
 &  & & & & \cr
$\frac{\Omega^2_0-\omega_0^2}{\omega_0^2}-4\xi$ &  $  -\frac{55}{6}\xi^2$    &      
    $ -\frac{71}{8}\xi^2 $    &  $-11\xi^2$            & $-32\xi^2$       &   $-12\xi^2$    \cr
&  & & & & \cr
\hline
&  & & & & \cr
 $ \frac{g_0-g_B}{g_B}$                                                      &   $ -\frac{4}{3}\xi$                     &   
    $-4\xi$                                         & $ -6\xi$                          &  $-6\xi$      &    $0$   \cr
&  & & & & \cr
\hline
&  & & & & \cr
 $x_0-1$                                                                               &  $\frac{1}{3}  \xi^2 $               &      
    $ \frac{1}{8}\xi^2$                  &  $\frac{2}{3}\xi^2$    &  $0$     &   $0$    \cr
&  & & & & \cr
\hline
 \end{tabular}
\end{center}
\caption{Comparison of the observables  obtained in the framework of various RG schemes. The expansion
in the dimensionless parameter $\xi= \frac{g_B\hbar}{16\omega_0^3}$ is truncated keeping the terms up to
the second order. 
\label{tab:comp}}
\end{table}

All the RG schemes reproduce the first-order corrections in accord with the RSPE. This is
because these corrections are one-loop corrections according to the loop expansion and
the essentially one-loop RG equations describe the quantum effects at the one-loop order exactly,
independently of the  approximation schemes additionally used.
We see that all the RG schemes have it   common that  the second-order corrections deepen
the energy levels, tend to decrease the quartic coupling  and result in a field-independent
wavefunction renormalization $x_0\ge 1$.  Nevertheless, the truncation of the
perturbation expansion causes a strong dependence of the quantitative results
on the RG scheme used.

\section{Conclusions}\label{sec:con}

The role of wavefunction renormalization and dependence of the observables on the
RG scheme in the second order perturbation expansion has been investigated on the 
example of the anharmonic oscillator. The functional CSi RG scheme with the imaginary mass  and the ISg RG with
the bare quartic coupling as control parameters have been applied to  the 1PI and 2PI
effective actions, respectively. The CSi RG  evolution equation derived for the 
$N$-dimensional $O(N)$ symmetric oscillator has been solved
by making a quartic Ansatz for the 1PI effective action including terms for field-dependent 
wavefunction renormalization, since that is the only way to generate field-independent
wavefunction renormalization  by means of a one-loop evolution equation. The evolution 
equation has been solved keeping  the terms of the Neumann-expansion of the inverse 
matrix in the right-hand side of the evolution equation only
up to the second order in the renormalized propagator and those up to the quartic ones in 
the oscillator coordinate. It has been shown that the $O(N)$ symmetric anharmonic oscillator
exhibits only a single phase, independently of its dimension $N$. Furthermore, the
inclusion of next-to-leading order of the gradient expansion, the wavefunction 
renormalization  does not modify qualitatively neither the IR scaling laws nor the phase 
structure. It has also been established that the effect of wavefunction renormalization 
decreases with increasing dimension $N$ of the oscillator according to a power-law.

The ISg RG has been applied to the
2PI effective action, in order to obtain a one-looplike RG evolution equation again. Our study
was restricted to the one-dimensional oscillator in that case. An approximation scheme  
based
on the E2PC,  going beyond the IMA, has been put forward that corresponds the expansion
 of the
2PI effective action into the powers of the free 2-particle propagator $\cG^{\mr{tree}}$.
The RG evolution equation for the 2PI effective action has been rewritten as a coupled set of
evolution equations for the 2PI 4-point and 8-point vertex functions
by keeping the terms up to those of the order
$\ord{(\cG^{\mr{tree}})^4}$ in the effective action and those of the order
 $\ord{(\cG^{\mr{tree}})^2}$
in the Neumann-expansion of the inverse of the full-particle propagator. 
The solutions for the vertex functions have been inserted into the Schwinger-Dyson 
equation for
the propagator of the ground state that has been solved by making an  Ansatz for the
latter assuming the dominance of a single pole. The advantages of that ISg RG method as 
compared
to the CSi RG scheme are that it enables one to read off the field-independent wavefunction
renormalization directly  and it provides an UV finite result for the ground-state energy. 

Finally we compared the results obtained for the ground-state energy,  the `mass'
(in field-theoric term),  the renormalized quartic coupling, and the field-independent
wavefunction renormalization in the various RG schemes (ISg RG, CSi RG, WH RG) and the
Rayleigh-Schr\"odinger perturbation expansion (RSPE)
restricting ourselves to the second-order perturbation expansion in powers of the bare quartic
coupling $g_B$. While the first-order contributions and the sign of the second-order ones
are reproduced by all of these approximations there has been found a remarkable 
scheme-dependence of the weights of the second-order terms.

\section*{Acknowledgements}

The authors thank Janos Polonyi for useful discussions.
Our work is supported by the project T\'AMOP 4.2.1-08/1-2008-003.
The project is implemented through the New Hungary Development Plan co-financed
by the European Social Fund, and the European Regional  
Development Fund.

\appendix

\section{CSi RG scheme for the $N$-dimensional anharmonic oscillator in real-time
 formalism}

\subsection{Neumann-expansion}\label{neumann}\label{a:nemex}

For later use let us introduce the matrices
\bea
  C_{\omega,\omega'}&=& {\bar C}_{(c,\omega),(c,\omega')}= C_{\omega',\omega},
\eea
and
\bea
  B_{\omega,-\omega}&=& -2  \int_{\omega_1}
[v(\omega_1,-\omega_1)+ v(\omega,-\omega)]
q_{a,\omega_1}q_{a,-\omega_1} 
=
  -2  \int_{\omega_1}\biggl(- \hf y( \omega_1^2+ \omega^2)  
+ \frac{g}{12} \biggr)q_{a,\omega_1}q_{a,-\omega_1} ,\nn
    C_{\omega,-\omega}&=&-4 \int_{\omega_1}
[v(\omega_1,-\omega)+v(-\omega_1,\omega) ]
q_{a,\omega_1}q_{a,-\omega_1}
=
-8 \int_{\omega_1}\biggl( -\hf \yb \omega_1\omega
   + \frac{1}{4} \Yb (\omega_1^2+\omega^2)
 + \frac{g}{24}
   \biggr)q_{a,\omega_1}q_{a,-\omega_1}. \nn
\eea
The first two terms in the expansion \eq{neuexp} are quadratic in the field variable $q_{a,\omega}$ and
can be expressed in terms of the loop integrals $I_{n,s}$ defined in Eq. 
\eq{csrealoop} in a straightforward manner,
\bea
  -\frac{ N\hbar}{2}\int_\omega G_\omega &=&-\frac{N}{2} I_{1,0},
\eea
\bea
\frac{\hbar}{2} \int_{\omega} G_\omega^2 A_{(a,\omega),(a,-\omega)}&=&
\frac{N\hbar}{2}  \int_{\omega} G_\omega^2 B_{\omega,-\omega}
  + \frac{\hbar}{2}  \int_{\omega} G_\omega^2 C_{\omega,-\omega}\nn
&=&
\hf [N\yb-(N+2)\Yb] I_{2,0} \int_{\omega_1}\omega_1^2 q_{a,\omega_1}q_{a,-\omega_1}  \nn
&&
+\hf  \biggl( [N\yb-(N+2)\Yb] I_{2,2}
 - \frac{(N+2) g}{6} I_{2,0}
\biggr)\int_{\omega_1}  q_{a,\omega_1}q_{a,-\omega_1}.
\eea
The third term of the series \eq{neuexp} can be rewritten as
\bea
\lefteqn{
-\frac{\hbar}{2}\int_{\omega,\omega'} G_\omega^2 A_{(c,\omega),(d,\omega')}G_{\omega'}A_{(d,-\omega'),(c, -\omega)}
}\nn
&=&
-\frac{\hbar}{2}\int_{\omega,\omega'} G_\omega^2 G_{\omega'}
[ N B_{\omega,\omega'}B_{-\omega,-\omega'}
+ 2B_{-\omega,-\omega'}C_{\omega,\omega'}
+{\bar C}_{(c,\omega),(d,\omega')}{\bar C}_{(c,-\omega),(d,-\omega')}]\nn
&=&
T_1+T_2+T_3,
\eea
where all terms are quartic in the Fourier-transform of the field variable,
\bea
 T_1&=&-\frac{N \hbar}{2}\int_{\omega,\omega'} G_\omega^2 G_{\omega'}
  B_{\omega,\omega'}B_{-\omega,-\omega'}\nn
&=&
- 2N\hbar\int_{\omega,\omega'} G_\omega^2 G_{\omega'}
\int_{\omega_1,\ldots,\omega_4} 
\delta_{\omega_1+\omega_2+\omega_3+\omega_4,0}
\delta_{\omega+\omega'+\omega_1+\omega_2,0}
\biggl( v(\omega_1,\omega_2)v(\omega_3,\omega_4)\nn
&&
+2v(\omega,\omega') v(\omega_1,\omega_2)
 +v^2(\omega,\omega')\biggr)
q_{a,\omega_1}q_{a,\omega_2} q_{b,\omega_3}q_{b,\omega_4},
\eea
\bea
T_2 &=&-\hbar \int_{\omega,\omega'} G_\omega^2 G_{\omega'} 
B_{-\omega,-\omega'}C_{\omega,\omega'} \nn
&=&
-8\hbar \int_{\omega,\omega',\omega_1,\omega_2,\omega_3,\omega_4}
G_\omega^2 G_{\omega'} \delta_{\omega_1+\omega_2+\omega_3+\omega_4,0}
\delta_{-\omega-\omega'+\omega_1+\omega_2,0}
\biggl( v(\omega_1,\omega_2) [v(\omega_3,\omega)
     + v(\omega_3,\omega')]\nn
&&
+v(\omega, \omega') [ v(\omega_3,\omega)+ v(\omega_3,\omega')] \biggr)
q_{a,\omega_1}q_{a,\omega_2} q_{b,\omega_3}q_{b,\omega_4} ,
\eea
\bea
T_3 &=& -\frac{\hbar}{2}\int_{\omega,\omega'} G_\omega^2 G_{\omega'}
{\bar C}_{(c,\omega),(d,\omega')}{\bar C}_{(c,-\omega),(d,-\omega')}  \nn
&=&
- 8\hbar \int_{\omega_1,\ldots,\omega_4,\omega,\omega'} G_\omega^2 G_{\omega'}
\delta_{\omega_1+\ldots+\omega_4,0}
\delta_{\omega_1+\omega_3+\omega+\omega',0}\nn
&&\times
[  v(\omega_1,\omega)v(\omega_2,-\omega)
+v(\omega_1,\omega)v(\omega_4,-\omega')
+v(\omega_3,\omega')v(\omega_2,-\omega)
+v(\omega_3,\omega')v(\omega_4,-\omega')]\nn
&&\times
q_{a,\omega_1}q_{a,\omega_2}q_{b,\omega_3}q_{b,\omega_4}.
\eea
Evaluating $T_{1}$, $T_2$, and $T_3$ we keep only the terms corresponding to the
interaction  vertices  included in the Ansatz \eq{ans} for the effective action in the 
truncated gradient expansion, ie. we keep self-interaction terms up to the quartic ones and 
quadratic gradient terms with wave-function renormalization truncated at the quadratic
field-dependent term. By making use of the Dirac-deltas expressing energy conservation in
the vertices, $G_{\omega'}$ can be changed to 
$G_{\omega+\alpha}$ with $\alpha=\pm(\omega_1+\omega_2)$. In order to reveal the
terms of the gradient expansion explicitly, we have to expand the
propagators with shifted frequency $G_{\omega+\alpha}$ in the Taylor series 
\eq{exprop} at $\alpha=0$. Furthermore, we also make use of the identity \eq{ident}, in
order to transform all gradient terms in one of those of the Ansatz \eq{ans}. A rather lengthy
but straightforward calculation yields finally,
\bea\label{t1}
T_1
&=&
- 2N\int_{\omega_1,\ldots,\omega_4} 
\delta_{\omega_1+\omega_2+\omega_3+\omega_4,0}
\biggl\{
 \frac{1}{4}y^2 I_{3,4}
- 2 \frac{g}{24}y I_{3,2} 
+4\frac{ g^2 }{24^2}I_{3,0}
\nn
&&
+ \biggl[
 2 y^2x^2 I_{5,6}
-16\frac{g}{24}yx^2 I_{5,4}
+32\frac{ g^2 }{24^2}x^2 I_{5,2}\nn
&&
-\frac{5}{2}x   y^2 I_{4,4}
+ 12\frac{g}{24}x yI_{4,2}
-8\frac{ g^2 }{24^2}x I_{4,0} \nn
&&
+\hf\yb^2 I_{3,2}-2\yb\Yb I_{3,2}+ \frac{3}{2}\Yb^2I_{3,2}
+ 4  \frac{g}{24} \Yb I_{3,0}
\biggr]\omega_1^2\nn
&&+ \biggl[ 
  2 y^2 x^2 I_{5,6}
   -16\frac{g}{24}y x^2 I_{5,4}
 +32\frac{ g^2 }{24^2}x^2 I_{5,2}\nn
&&
-\frac{5}{2} x   y^2 I_{4,4}
+12\frac{g}{24}y x I_{4,2}
-8\frac{ g^2 }{24^2}xI_{4,0}\nn
&&
-\yb\Yb I_{3,2}+\Yb^2 I_{3,2}
+2 \frac{g}{24}\yb I_{3,0}
+  2\frac{g}{24} \Yb I_{3,0}
\biggr]\omega_1\omega_2
\biggr\}q_{a,\omega_1}q_{a,\omega_2} q_{b,\omega_3}q_{b,\omega_4},
\eea
\bea\label{t2}
T_2
&=&
-8 \int_{\omega_1,\omega_2,\omega_3,\omega_4} 
 \delta_{\omega_1+\omega_2+\omega_3+\omega_4,0}\nn
&&\times
\biggl\{
-\frac{1}{4}y \Yb  I_{3,4}
+  2  \frac{g}{24}\Yb  I_{3,2} 
-  \frac{g}{24}\yb  I_{3,2}
+ 4\frac{g^2}{24^2}I_{3,0}
\nn
&&
+\biggl[
        -   2 x^2 y \Yb I_{5,6}
-8\frac{g}{24}x^2 \yb I_{5,4} 
+16 \frac{g}{24}x^2  \Yb I_{5,4}
+32\frac{g^2}{24^2}  x^2  I_{5,2} \nn
&&
 +\frac{5}{2}x y \Yb I_{4,4}
+6\frac{g}{24}       x \yb I_{4,2} 
 -  12\frac{g}{24}       x \Yb I_{4,2}
-8\frac{g^2}{24^2}       x  I_{4,0}\nn
&&
+\frac{1}{4} \yb^2 I_{3,2}
     - 5\frac{1}{4}\yb \Yb I_{3,2}
+ \frac{3}{2}\Yb^2 I_{3,2} 
+ 4\frac{g}{24} \Yb I_{3,0}
-\frac{g}{24} \yb  I_{3,0} 
\biggr]\omega_1^2\nn
&&+
\biggl[
-2x^2  y \Yb  I_{5,6}  
-8\frac{g}{24}    x^2 \yb I_{5,4}
  +16\frac{g}{24}    x^2 \Yb I_{5,4} 
+32\frac{g^2}{24^2}    x^2 I_{5,2}
\nn
&&
+\frac{5}{2}x y \Yb I_{4,4}
+6\frac{g}{24} x \yb I_{4,2}
-  12\frac{g}{24} x   \Yb  I_{4,2} 
-8\frac{g^2}{24^2} x  I_{4,0}
\nn
&&
+ \frac{1}{4} \yb^2 I_{3,2}
    - 3 \frac{1}{4} \yb  \Yb I_{3,2}
       +  \Yb^2   I_{3,2}
+ 2\frac{g}{24}   \Yb_{\mu^2} I_{3,0}
  \biggr]\omega_1\omega_2
\biggr\}\nn
&&\times
q_{a,\omega_1}q_{a,\omega_2} q_{b,\omega_3}q_{b,\omega_4},
\eea
and
\bea\label{t3}
T_3
&=&
- 8\hbar \int_{\omega_1,\ldots,\omega_4,\omega} 
\delta_{\omega_1+\ldots+\omega_4,0}\nn
&&\times
\biggl[  \frac{\Yb^2}{4}   I_{3,4}
+ \frac{g\Yb}{12}I_{3,2} 
+ 4\frac{g^2}{24^2} I_{3,0}\nn
&&
+\biggl(
\Yb^2 x^2I_{5,6}
+ 8\frac{g\Yb}{24}x^2 I_{5,4}
+16\frac{g^2}{24^2}  x^2I_{5,2}
\nn
&&
-20\frac{\Yb^2}{16}xI_{4,4}
- 6\frac{g\Yb}{24} x I_{4,2}
-4\frac{g^2}{24^2} x I_{4,0}
\nn
&&
+\Yb^2I_{3,2}
-\frac{1}{4} \yb^2I_{3,2}
- \frac{\yb \Yb}{4} I_{3,2}
- \frac{g \yb}{24}  I_{3,0} 
+ 3 \frac{g\Yb}{ 24}I_{3,0}
\biggr)\omega_1^2   \nn
&&
+\biggl(- \Yb^2 x^2 I_{5,6}
- 8\frac{g\Yb}{24}x^2 I_{5,4}
- 16\frac{g^2}{24^2}  x^2 I_{5,2} \nn
&&
+5\frac{\Yb^2}{4}x I_{4,4}
+6\frac{g\Yb}{24} x I_{4,2}
+4\frac{g^2}{24^2} x I_{4,0}\nn
&&
-\frac{3}{4} \yb^2I_{3,2}
- \hf \Yb^2I_{3,2}
+ \frac{\yb \Yb}{4}I_{3,2}
+ \frac{g \yb}{24}  I_{3,0} 
- \frac{g\Yb}{ 24}I_{3,0}
\biggr)\omega_1\omega_2
\biggr]\nn
&&\times
q_{a,\omega_1}q_{a,\omega_2}q_{b,\omega_3}q_{b,\omega_4}.
\eea

\subsection{Loop integrals}\label{a:loopCS2}

In order to get the explicit forms of the RG equations \eq{ga}-\eq{Yb}  one has  to perform 
the loop integrals \eq{csrealoop},
\bea
I_{n,s}&\sim&\int_{-\infty}^\infty \frac{d\omega}{2\pi} G_\omega^n\omega^s
= \int_{-\infty}^\infty \frac{d\omega}{2\pi x^n}\frac{ \omega^s}{ (\omega-a)^n
(\omega+b)^n}\biggr|_{a=b=\sqrt{(\Omega^2\v{-}i\mu^2)/x} }.
\eea
For the limit $\Lambda\to \infty$ and for $s-2n<-1$,  like in our case, this can be done 
analytically. Since ${\mr{Re~}}a>0$, ${\mr{Im~}}a>0$, we can close the path of
integration with a half-circle of infinite radius on the upper half of the complex 
$\omega$-plane encircling the  multiple pole at $\omega=a$ and apply the residuum 
theorem,
\bea
  \frac{1}{2\pi} \oint \frac{f(z)}{(z-a)^n}&=& \frac{i}{(n-1)!}\frac{d^{n-1}}{da^{n-1}}
f(a),~~~~f(a)=\frac{a^s}{(a+b)^n}.
\eea
Thus one finds 
\bea\label{loopi}
 && I_{1,0}=  \frac{i }{2x^\hf  (\Omega^2 \v{-}i\mu^2)^{\frac{1}{2}}},
~~~~
 I_{n,s}=
\frac{i  [s-(2n-3)]\cdot (s-1) }{(n-1)!2^nx^{\frac{s+1}{2} }
 (\Omega\v{-}i\mu^2)^{n-\frac{s+1}{2}}} ~~{\mbox{for~~}}n>1, 
 \eea
which yield the relations
\bea
&&I_{2,2}= \frac{1}{2x}I_{1,0},~~
I_{3,2}= \frac{1}{4x}I_{2,0},~~
I_{3,4} =\frac{3}{8x^2}I_{1,0},~~
I_{4,2}=\frac{1}{6x}I_{3,0}, ~~ 
I_{4,4}=\frac{1}{8x^2}I_{2,0},\nn
&&I_{5,2}=\frac{1}{8x}I_{4,0},~~
I_{5,4}=\frac{1}{16x^2} I_{3,0},~~
I_{5,6}=\frac{5}{64x^3}I_{2,0}
\eea
with
\bea
 &&I_{2,0}= \frac{-i    }{4x^{\frac{1}{2} }
 (\Omega\v{-}i\mu^2)^{\frac{3}{2}}} ,~~
I_{3,0}=\frac{3i   }{16 x^{\frac{1}{2} }
 (\Omega\v{-}i\mu^2)^{\frac{5}{2}}},~~
I_{4,0}=\frac{-5 i   }{32 x^{\frac{1}{2} }
 (\Omega\v{-}i\mu^2)^{\frac{7}{2}}}  .
\eea

\subsection{Perturbation expansion}\label{a:cspe}

Here we outline the solution of the evolution equations \eq{gan1}-\eq{upsi} for the $N=1$
dimensional oscillator in the second order of the perturbation expansion in the bare coupling $g_B$
after the analytic continuation $-i\mu^2\to \lambda$ to the Euclidean space.
Making use of the result of the IMA, one can write the perturbation expansion of the couplings as,
\bea
  \gamma&=& \delta+g_B\alpha+g_B^2\beta+\ldots,\nn
\delta \Omega^2 &=& g_B\rho+g_B^2\sigma+\ldots,\nn
x-1 &=& g_B^2\zeta+\ldots,\nn
\Upsilon &=& g_B^2\eta,\nn
g&=& g_B+g_B^2\nu
\eea
with $\delta \Omega^2=\Omega^2- \omega_0^2$, and using $x -1\sim \ord{\Upsilon}\sim \ord{g_B^2}$.
For the sake of simplicity, we shall set $\hbar=1$ in this section.
In order to make explicit the dependence on the bare coupling $g_B$ in the right-hand 
sides of Eqs. \eq{gan1}-\eq{upsi}, one has to expand the denominators of the loop integrals
given by Eq. \eq{loopi}, 
\bea
\frac{1}{(\Omega^2+\lambda)^a}&=&
   \frac{1}{(\omega_0^2+\lambda)^a}
 - a\frac{g_B\rho }{(\omega_0^2+\lambda)^{a+1} } 
+ g_B^2 \biggl( 
\hf a(a+1) \frac{\rho^2}{(\omega_0^2+\lambda)^{a+2} }
    - \frac{a\sigma}{  (\omega_0^2+\lambda)^{a+1} } \biggr),\nn
 x^a &\approx & 1 + a g_B^2 \zeta+\ldots,
\eea
where the relation
\bea
\frac{1}{(1+x)^a}&\approx& 1 -a x + \hf a (a+1) x^2+\ldots
\eea
has been used. The comparison of the coefficients of the corresponding powers of $g_B$
in both sides of the Eqs. \eq{gan1}-\eq{upsi} yields the following set of coupled ordinary
first-order differential equations for the running couplings,
\bea\label{del}
\partial_{\lambda} \delta
&=& 
   \frac{1}{4(\omega_0^2+\lambda)^\frac{1}{2}},
\eea
\bea\label{alp}
\partial_{\lambda}  \alpha&=& 
 -\frac{1}{8}\frac{\rho }{(\omega_0^2+\lambda)^{\frac{3}{2}} }, 
\eea
\bea\label{bet}
\partial_{\lambda} \beta
&=& 
\frac{3}{32} \frac{\rho^2}{(\omega_0^2+\lambda)^\frac{5}{2} }
    + \frac{1}{8}\frac{\sigma}{  (\omega_0^2+\lambda)^\frac{3}{2} } 
    + \frac{1}{8} \frac{\zeta}{(\omega_0^2+\lambda)^\frac{1}{2}},
\eea
\bea\label{zet}
\partial_{\lambda}\zeta &=& 
 - \frac{\eta}{4(\omega_0^2+\lambda)^\frac{3}{2}},
\eea
\bea\label{eta}
\partial_{\lambda}\eta &=&
-  \frac{5}{64(\omega_0^2+\lambda)^\frac{7}{2}},
\eea
\bea\label{rho}
\partial_{\lambda}\rho &=&
 - \frac{1 }{8(\omega_0^2+\lambda)^\frac{3}{2}},
\eea
\bea\label{sig}
\partial_{\lambda}\sigma &=&
-\frac{\eta}{4 (\omega_0^2+\lambda)^\hf}
 -\frac{\nu }{8(\omega_0^2+\lambda)^\frac{3}{2}}
 + \frac{3}{16}\frac{\rho }{(\omega_0^2+\lambda)^{\frac{5}{2}} } ,
\eea
\bea\label{nu}
\partial_{\lambda}\nu &=&
  \frac{9  }{ 16 (\omega_0^2+\lambda)^\frac{5}{2}}.
\eea
The solutions belonging to the initial conditions $\delta_B=-\hf (\omega_0^2+\lambda_B)^\hf$,
$\alpha_B=\beta_B=\zeta_B=\eta_B=\rho_B=\sigma_B=\nu_B=0$ at $\lambda_B\to \infty$ and
their IR limits $\lambda\to 0$ have been found analytically,
\bea
&&\delta=\hf (\omega_0^2+\lambda)^\hf \to \hf \omega_0,\nn
&&\eta=\frac{1}{32 (\omega_0^2+\lambda)^\frac{5}{2} }\to
\frac{1}{32\omega_0^5 },\nn
&&\rho=\frac{1}{4(\omega_0^2+\lambda)^\hf}
\to\frac{1}{4\omega_0},\nn
&&\nu=-\frac{3}{8(\omega_0^2+\lambda)^\frac{3}{2} }
\to -\frac{3}{8\omega_0^3},\nn
&&\sigma=- \frac{11}{256 (\omega_0^2+\lambda)^2}
\to  -\frac{11}{256\omega_0^4},\nn
&& \alpha= \frac{1}{32(\omega_0^2+\lambda)}\to \frac{1}{32\omega_0^2},\nn
\nn
&&\zeta=\frac{1}{3\cdot 128 (\omega_0^2+\lambda)^3}
\to \frac{1}{3\cdot 128\omega_0^6} ,\nn
&&\beta= - \frac{71}{3\cdot 5\cdot 1024(\omega_0^2+\lambda)^\frac{5}{2}}
\to -\frac{71}{3\cdot 5\cdot  1024 \omega_0^5}.
\eea
Then one obtains for the energy of the ground state
\be
  E_0=\gamma_0
= \hf \omega_0  \biggl[ 1 +\xi     - \frac{71}{30}\xi^2 \biggr],
\ee
for the frequency parameter
\be
\Omega^2_0=\omega_0^2(1 + 4\xi-11\xi^2),
\ee
the renormalized quartic coupling
\be
g_0= g_B(1 -6\xi),
\ee
and the couplings for the wavefunction renormalization
\be
  x_0= 1 +\frac{2}{3}\xi^2,~~
 \hbar\Upsilon_0= 8\omega_0\xi^2,
\ee 
where we reestablished the powers of $\hbar$.

\subsection{Localization}\label{a:csloc}

The localization of the ground state wavefunction can be characterized by its momenta.
We can make an estimate of the dependence of the variance of the coordinate operator
on the dimension $N$ of the oscillator.
Let us write for it with the help of the Lehmann-expansion of the propagator,
\bea
  \langle 0| \u{q}^2 |0\rangle &=& \sum_{a=1}^N  \langle 0| q_a^2 |0\rangle
= \lim_{t\to 0^+} \sum_{a=1}^N \sum_{n,\alpha} \langle 0| q_a(t)|n,\alpha\rangle\langle 
0,\alpha|q_a(0)|
0\rangle \nn
&=& \lim_{t\to 0^+}   \sum_{a=1}^N  \sum_{n,\alpha} e^{i(E_0-E_n)t/\hbar} | \langle 0| q_a|n, \alpha\rangle |^2\nn
&=& \sum_{a=1}^N \sum_{n,\alpha} \lim_{t\to 0^+} \int_\omega e^{i\omega t}
\frac{ 2(E_n-E_0)/\hbar}{\omega^2+ [(E_n-E_0)/\hbar]^2}  | \langle 0| q_a|n,\alpha\rangle |^2
\eea
where $H|n, \alpha\rangle=E_n|n,\alpha\rangle$ for the exact eigenstates $|n, \alpha\rangle$ and
energy levels $E_n$ of the Hamiltonian, $|0\rangle $ and $E_0$ stand for the ground state and its energy,
$\alpha$ are additional quantum numbers for counting states belonging to the same degenerate energy level.
Operators without and with the time argument stand for those in the Schr\"odinger and the
Heisenberg representations, respectively. Making use of the dominance of the pole of the 
integrand corresponding to the first excited state \cite{Aoki2002}, one finds the estimate
\bea
  \langle 0|\u{q}^2|0\rangle&\approx & \lim_{t\to 0^+}\int_\omega
\frac{e^{i\omega t}}{\omega^2 +(\Omega^2/x)} \sum_{a=1}^N
\frac{1}{\hbar}\sum_{n,\alpha} 2(E_n-E_0) | \langle 0| q_a|n,\alpha\rangle |^2\nn
&=&
\frac{x^\hf}{2\Omega}\sum_{a=1}^N
\frac{1}{\hbar}\sum_{n,\alpha} 2(E_n-E_0) | \langle 0| q_a|n,\alpha\rangle |^2.
\eea
Making use of the Thomas-Reiche-Kuhn sum rule \cite{Thom1925}, valid for any self-adjoint operator $A$,
\be
 \hf \langle 0|[ [M,H],M]|0\rangle = \sum_{n, \alpha} (E_n-E_0) |\langle 0| M|n, \alpha
  \rangle |^2
\ee
and the form of the renormalized Hamilton-operator 
$H=\hf xp^2+ V(q)$ where we neglected the less important higher-momentum terms
in respect to the small values of $\yb$ and $\Yb$, we get
\bea
 \sum_{n, \alpha} (E_n-E_0) |\langle 0| q_a|n, \alpha
  \rangle |^2 &=&
\hf \langle 0| [[q_a,H],q_a]|0\rangle 
= \frac{x}{4} \langle 0| [[q_a,p_a^2],q_a] |0\rangle = \hf x \hbar^2
\eea
where there is no summation over the index $a$. With the help of this relation we obtain
\be\label{qvar}
   \langle 0 |\u{q}^2|0\rangle \approx \frac{x^\hf}{2\Omega} \frac{2}{\hbar}
\sum_{a=1}^N \hf x \hbar^2 =\frac{Nx^{3/2} \hbar}{2\Omega}.
\ee

\section{WH RG flow of the Wilsonian action in Euclidean space}\label{a:wheuc}

\subsection{Blocking via mode-by-mode integration}

The WH RG method \cite{Wegn1973} has been applied to simple quantum mechanical systems with success in
the literature \cite{Aoki2002}. The simplicity of the $0+1$ dimensional model of the single quantized
anharmonic oscillator enables one to perform the blocking by integrating out the high-frequency quantum 
fluctuations mode-by-mode. Let us prescribe periodic boundary conditions in time, $q_{t=0}=q_{t=T}$, $\xi_{t=0}=
\xi_{t=T}$ both for the coordinate $q_t$ and the mode $\xi_t$ integrated out in a single
blocking step, where $T$ stand for the time interval the action is integrated over and taken to infinity at the end.
The Matsubara frequencies are given as $\omega_n=n2\pi/T$, the gliding cut-off is at
$\omega_N=N2\pi/T$. The periodic coordinate can be Fourier expanded as
\be
  q_t = \frac{a_0}{2} +\sum_{n=1}^{N_\Lambda} a_n\cos(\omega_n t) 
+\sum_{n=1}^{N_\Lambda} b_n\sin(\omega_n t) ={\bar q}+\eta_t
\ee
where $\omega_{N_\Lambda}=N_\Lambda 2\pi/T=\Lambda$ is the UV cut-off. 
Integrating out all the quantum fluctuations $\eta_t$ around the constant mode ${\bar q}=\hf a_0$
leads to the effective action $S_0({\bar q})$ given by
\bea
   e^{-\frac{1}{\hbar} S_0({\bar q})}
&=& \int{\cal D}\eta  e^{-\frac{1}{\hbar} S_\Lambda [{\bar q}+\eta]  }\nn   
 &\sim& \biggl( \prod_{n=1}^{N_\Lambda}
\int_{-\infty}^\infty da_n \int_{-\infty}^\infty db_n \biggr) 
   e^{-\frac{1}{\hbar} S_\Lambda [{\bar q}+\eta] }   
\eea
up to an irrelevant constant being independent of ${\bar q}$ and the potential  (see
 \cite{Fey72}).
The blocking relation for integrating out the single mode with frequency  $\omega_N$, ie. the mode
$\xi_t = a_N \cos(\omega_N t) + b_N \sin(\omega_N t)$
 is given by
\bea
  e^{-\frac{1}{\hbar} S_{N-1}[{\bar q}]}&= & \int {\cal D}\xi
 e^{-\frac{1}{\hbar} S_{N}[{\bar q}+\xi]}
\sim \int_{-\infty}^\infty da_N  \int_{-\infty}^\infty db_N e^{-\frac{1}{\hbar} S_{N}[{\bar q}+\xi]}
\eea  
in the LPA. Here the LPA enables one to replace the time-dependent background  by the constant mode ${\bar q}$.
The kinetic term of the action gives
 \bea
  \int_0^T dt \hf ({\dot q}_0+{\dot \xi}_t)^2 &=&
  \frac{T}{4} \omega_N^2 [a_N^2+ b_N^2] .
\eea
With the notation $a_{\sigma=1}=a_N$, $a_{\sigma=-1}=b_N$, we can write for the blocked potential
 \bea
{\cal U}_N[{\bar q}+\xi]&=&\int_t U_N({\bar q}+\xi_t)=
   TU_N({\bar q}) + \hf \sum_{\sigma, \sigma'=\pm 1} a_{\sigma } a_{\sigma' } 
   {\cal U}^{(2)}_{N~\sigma, \sigma'}({\bar q})+ \ord{\xi^3},
\eea
where
\bea
 {\cal U}^{(2)}_{N~\sigma, \sigma'}({\bar q})&=&
\fdd{}{ a_\sigma'}{a_\sigma} \int_t U_N({\bar q}+\xi_t)
\biggr|_{\xi=0}
=
 T \hf  U^{(2)}_N({\bar q}) [ \delta_{\sigma,1}\delta_{\sigma',1} + \delta_{\sigma,-1}\delta_{\sigma',-1}],
\eea
so that one gets
 \bea
{\cal U}_N[{\bar q}+\xi]&=&
  TU_N({\bar q}) +T \frac{1}{4}U^{(2)}_N({\bar q})[a_N^2+b_N^2] + \ord{\xi^3}.
\eea
The non-Gaussian higher-order terms are suppressed with some powers of $1/T$ in the
path-integral and can be neglected.  Then the Gaussian path integral can be performed,
\bea
 e^{-\frac{1}{\hbar} TU_{N-1}({\bar q}) } &\sim &
  e^{-\frac{1}{\hbar}  TU_N({\bar q}) } \int_{-\infty}^\infty da_N\int_{-\infty}^\infty db_N
e^{-\frac{1}{\hbar} T \frac{1}{4}  [ \omega_N^2+ U^{(2)}_N({\bar q})]
[a_N^2+b_N^2]  } \nn
&\sim &
  e^{-\frac{1}{\hbar}  TU_N({\bar q}) } \biggl( \sqrt{ 
   \frac{8\pi\hbar}{ T[ \omega_N^2+ U^{(2)}_N({\bar q})]}     }\biggr)^2,
\eea
yielding the finite difference equation
\bea
 \frac{ U_N({\bar q})-U_{N-1}({\bar q}) }{2\pi/T} =
\frac{ U_{\omega_N}({\bar q}-U_{\omega_N-\Delta \omega}({\bar q}) }{ \Delta \omega}
&=& - \frac{\hbar/T}{2\pi/T} \ln [ \omega_N^2+ U^{(2)}_N({\bar q})]
\eea
for the blocked potential at the scale $\omega_N$. 
In the limit $T\to \infty$, ie. $\Delta \omega\to 0$ one finds the WH RG equation
\bea\label{wheq}
 \partial_\omega U_\omega({\bar q})&=& - \hbar \alpha \ln[ \omega^2 +
    U^{(2)}_\omega({\bar q})],~~~~\alpha=\frac{1}{2\pi}
\eea
for the blocked potential $U_\omega$, where the discrete gliding frequency cut-off
$\omega_N$ has been changed to the continuous one, denoted by $\omega$.  

\subsection{Approximation schemes}

\subsubsection{IMA}

In the IMA one has to replace the blocked potential in the right-hand side by the bare one.
Then one can perform the integration over the scale $\omega$ in a straightforward manner,
\bea
  U_\omega^{\rm{IMA}}({\bar q})&=&U_\Lambda({\bar q}) +\hbar \alpha
   \int_\omega^\Lambda dk \ln [k^2+ a^2],~~~~a^2=U^{(2)}_\Lambda({\bar q}).
\eea
The bare potential is convex at least in the neighbourhood of its minimum (minima) at some $q_0$.
Then for ${\bar q}$ sufficiently close to $q_0$, one can make use of the integral
\bea
  \int_\omega^\Lambda dk \ln [k^2+ a^2]&=&
    \biggl[k \ln  [k^2+ a^2] - 2k
+ 2a{\mbox{~arc~tan~}}\frac{k}{a}\biggr]_\omega^\Lambda 
\stackrel{\Lambda\to \infty,~\omega\to 0}{\longrightarrow}  
  2\Lambda (\ln \Lambda -1) + a\pi 
\eea
and find for the effective potential  
\bea
   U_0^{\rm{IMA}}({\bar q})&=& U_\Lambda({\bar q}) + \hbar   \pi \sqrt{ U^{(2)}_\Lambda({\bar q})} .
\eea
This expression is valid in the neighbourhood of the minimum $q_0$ of the bare potential,
and the UV divergent piece is removed by the renormalization condition that the effective
potential of a free particle  $(U_\Lambda({\bar q})=0, ~a=0)$ were identically vanishing.

The minimum at $q_0$ of the bare potential would be displaced to $q_0^{IMA}=q_0+\delta q$ with 
\be
   \delta q= -\frac{\hbar \pi U^{(3)}_\Lambda (q_0) }{2 [ U^{(2)}_\Lambda (q_0)]^{3/2} }
\ee
so that for a potential with the symmetry ${\bar q}\to -{\bar q}$ and its minimum at $q_0=0$
the minimum is not displaced in the one-loop order. Then one finds for the ground state energy 
\be
   E_0^{{\rm{IMA}}}= \hbar \alpha \pi  \sqrt{ U^{(2)}_\Lambda(0)} 
\ee
in the IMA, ie. in the one-loop approximation. For the free motion of a particle the bare potential
vanishes and it remains vanishing after the blocking as well.
For the case of a linear harmonic oscillator this reproduces the correct ground state energy
$E_0^{{\rm{IMA,~ lho.}}}=\hf \hbar \omega_0$.
For an arbitrary polynomial bare potential $U_\Lambda ({\bar q})=
\hf \omega_0^2 {\bar q}^2 + \frac{1}{24} g_B {\bar q}^4+\ldots$
we get
\bea
  U_0^{\rm{IMA,~ aho.}}({\bar q}) &=&U_\Lambda({\bar q})+
 \hf \hbar \omega_0( 1 
+ \frac{g_B{\bar q}^2}{4\omega_0^2}  +\ldots ) =
  \hf \hbar \omega_0 +\hf \biggl( \omega_0^2
+  \frac{ g_B \hbar }{4\omega_0}\biggr)  {\bar q}^2 
+\frac{1}{24} g_B {\bar q}^4+\ldots
\eea
This means that the energy of the ground state of the anharmonic oscillator remains the same as that of
the harmonic oscillator, but the excitation energy of the first excited state 
is increased with increasing coupling $g_B$,
\be
\Delta E^{\mr{IMA}}=  \hbar \sqrt{ \omega_0^2 + \frac{ g_B \hbar }{4\omega_0} }
 \approx \hbar \omega_0\biggl( 1 + \frac{ g_B \hbar }{8\omega_0^3}+\ord{g_B^2} \biggr),
\ee
in the IMA. The same result has been obtained in the CSi RG scheme in the IMA.

\subsubsection{Perturbation expansion in powers of the bare coupling $g_B$}\label{whpe}

Let us introduce the blocked interaction potential $ V_\omega ({\bar q})$ via
\be
  U_\omega({\bar q})= u_\omega + \hf \omega_0^2 {\bar q}^2 + V_\omega ({\bar q}),
\ee
for which the WH-RG equation \eq{wheq} can be rewritten as 
\bea\label{whrgper}
 \partial_\omega u_\omega+\partial_\omega V_\omega ({\bar q})&=&
  -\hbar \alpha\ln [ \omega^2 + \omega_0^2+ V_\omega''({\bar q})]
=
 - \hbar \alpha \ln(\omega^2+\omega_0^2)
  - \hbar \alpha\sum_{n=1}^\infty \frac{-(-1)^n}{n}
[G_{0~\omega}V_\omega''({\bar q})]^n .
\eea
In the last equation we performed the perturbation expansion of the right hand side into
the interaction, $G_0$ denotes the propagator of the harmonic oscillator in the ground state.
Further on let us Taylor-expand the blocked interaction potential into the powers of ${\bar q}$
and truncate it at the quartic terms,
\bea
 V_\omega({\bar q})&=& g_B\alpha_\omega + g_B^2\beta_\omega
   + \hf (g_B \rho_\omega+ g_B^2 \sigma_\omega){\bar q}^2 +
   \frac{1}{24} (g_B+g_B^2 \nu_\omega) {\bar q}^4+\ldots,
\eea
and Taylor-expand each running coupling into the
powers of the bare coupling $g_B$ and truncate their perturbation serii at the second order
terms. Inserting this ansatz into Eq. \eq{whrgper} and keeping all the terms up to the
order ${\bar q}^4$ and $g_B^2$, one finds the following set of coupled ordinary
differential equations for the RG flow of the various couplings $(\dot ~~=d/d\omega)$,
\bea
{\dot u}_\omega &=& -\hbar \alpha \ln (\omega^2+\omega_0^2),\nn
{\dot \alpha}_\omega &=& 
-\hbar \alpha\frac{\rho_\omega}{\omega^2+\omega_0^2},\nn
{\dot \beta}_\omega &=&
 -\hbar \alpha\frac{\sigma_\omega}{\omega^2+\omega_0^2} 
+ \hf \hbar \alpha \frac{\rho_\omega^2}{(\omega^2+\omega_0^2)^2},
\nn
{\dot \rho}_\omega&=&
  -\hbar \alpha\frac{1}{\omega^2+\omega_0^2},\nn
{\dot \sigma}_\omega &=& 
 -\hbar\alpha\frac{\nu_\omega}{\omega^2+\omega_0^2} 
+ \hbar\alpha\frac{\rho_\omega}{(\omega^2+\omega_0^2)^2 },\nn
{\dot \nu}_\omega&=& 
3\hbar \alpha\frac{1}{(\omega^2+\omega_0^2)^2}
\eea
with the initial conditions $\alpha_B=\beta_B=\rho_B=\sigma_B=\nu_B=0$ at the UV cut-off $\omega=\Lambda$.
Here the equations for $u_\omega,~\rho_\omega,~\nu_\omega, ~\alpha_\omega$ can be integrated 
straightforwardly,
\bea
u_\omega&=& 
u_\Lambda +
\hbar \alpha\int_\omega^\Lambda dk\ln (k^2+\omega_0^2)
\stackrel{\Lambda\to \infty,~\omega\to 0}{\longrightarrow} 
 {\mbox{~UV~div.~terms~}}+ \hf \hbar \omega_0,
\eea
\bea
\rho_\omega&=& \frac{\hbar}{2\pi}\int_\omega^\Lambda \frac{dk}{k^2+\omega_0^2}
= \frac{\hbar}{2\pi\omega_0}\biggl[ {\mbox{~arc~tan~}}\frac{k}{\omega_0}
\biggr]^\Lambda_\omega 
\stackrel{\Lambda\to \infty,~ \omega\to 0}{\longrightarrow}
\frac{\hbar}{4\omega_0}, 
\eea
\bea
\nu_\omega&=& 
-\frac{3\hbar}{2\pi}\int_\omega^\Lambda \frac{dk}{(k^2+\omega_0^2)^2 } =  -\frac{3\hbar}{2\pi}
\biggl[ \frac{k}{2\omega_0^2(k^2+\omega_0^2) }
  + \frac{1}{2\omega_0^3}{\mbox{~arc~tan~}}\frac{k}{\omega_0} 
\biggr]^\Lambda_\omega  
\stackrel{\Lambda\to \infty,~\omega\to 0}{\longrightarrow}
-\frac{3\hbar}{8\omega_0^3},
\eea
\bea
\alpha_\omega &=& \frac{\hbar^2\alpha^2}{\omega_0}\int_\omega^\Lambda dk
\frac{ {\mbox{~arc~tan~}}\frac{\Lambda}{\omega_0}-
  {\mbox{~arc~tan~}}\frac{k}{\omega_0}     }{k^2+\omega_0^2}
\nn
&=&
 \frac{\hbar^2\alpha^2}{\omega_0}\biggl[
\frac{1}{\omega_0} {\mbox{~arc~tan~}}\frac{\Lambda}{\omega_0}
\biggl( {\mbox{~arc~tan~}}\frac{k}{\omega_0} \biggr)_\omega^\Lambda 
-\frac{1}{2\omega_0} \int_\omega^\Lambda dk~ 2 \biggl( {\mbox{~arc~tan~}}\frac{k}{\omega_0}
 \biggr)^\prime {\mbox{~arc~tan~}}\frac{k}{\omega_0}
\biggr]\nn
&=&
 \frac{\hbar^2\alpha^2}{\omega_0}\biggl[
\frac{1}{\omega_0} {\mbox{~arc~tan~}}\frac{\Lambda}{\omega_0}
\biggl( {\mbox{~arc~tan~}}\frac{k}{\omega_0} \biggr)_\omega^\Lambda 
-\frac{1}{2\omega_0} \biggl({\mbox{~arc~tan}}^2~\frac{k}{\omega_0}  \biggr)^\Lambda_\omega\biggr]
\stackrel{\Lambda\to \infty,~\omega\to 0}{\longrightarrow} 
 \frac{\hbar^2}{32\omega_0^2} .
\eea
The rest of the equations cannot be integrated straightforwardly, but their solutions can be
approximated by inserting the IR limits of the couplings $\rho_\omega$, $\nu_\omega$
into their right hand sides,
\bea
\sigma_\omega&\approx & 
   -\frac{3\hbar^2}{16\pi\omega_0^3}\int_\omega^\Lambda \frac{dk}{k^2+\omega_0^2} 
 - \frac{\hbar^2}{8\pi\omega_0}\int_\omega^\Lambda \frac{dk}{(k^2+\omega_0^2)^2} 
\stackrel{\Lambda\to \infty,~k\to 0}{\longrightarrow} 
- \frac{\hbar^2}{8\omega_0^4},
\eea
\bea
 \beta_\omega&\approx & \frac{\hbar}{2\pi}\frac{-\hbar^2}{8\omega_0^4}
\int_\omega^\Lambda \frac{dk}{k^2+\omega_0^2} 
-\frac{\hbar}{4\pi}\frac{\hbar^2}{16\omega_0^2}\int_\omega^\Lambda 
\frac{dk}{(k^2+\omega_0^2)^2}
\stackrel{\Lambda\to \infty,~k\to 0}{\longrightarrow} 
-\frac{9\hbar^3}{256\omega_0^5}.
\eea
\normalsize
Finally one finds
\bea
 E_0&=& u_0+g_B\alpha_0+g_B^2\beta_0
= \hf \hbar\omega_0 ( 1 +\xi - 18\xi^2+\ldots ),\nn
\Omega^2_0&=&\omega_0^2 + g_B \rho_0+ g_B^2 \sigma_0
= \omega_0^2 (1+ 4\xi - 32\xi^2+\ldots ),\nn
g_0&=&g_B+g_B^2 \nu_0= g_B( 1- 6\xi+\ldots )
\eea
in terms of the dimensionless parameter $\xi=g_B\hbar/(16\omega_0^3)$.
One should mention that the gradient expansion cannot be improved going beyond the 
LPA in the framework of the WH RG scheme.

\end{document}